\documentclass{article}

\usepackage{microtype}
\usepackage{graphicx}
\usepackage{subfigure}
\usepackage{booktabs} 

\usepackage{hyperref}

\usepackage[accepted]{icml2023}

\usepackage{amsmath}
\usepackage{amssymb}
\usepackage{mathtools}
\usepackage{amsthm}

\DeclareUnicodeCharacter{2064}{}

\usepackage[capitalize,noabbrev]{cleveref}

\theoremstyle{plain}

\theoremstyle{definition}

\theoremstyle{remark}

\usepackage[textsize=tiny]{todonotes}

\begin{document}
\icmltitlerunning{Large Language Models in Drug Discovery and Development}
\twocolumn[
\icmltitle{Large Language Models in Drug Discovery and Development: \\ From Disease Mechanisms to Clinical Trials}


\icmlsetsymbol{equal}{*}

\begin{icmlauthorlist}
\icmlauthor{Yizhen Zheng$^{1}$}{equal}
\icmlauthor{Huan Yee Koh$^{1, 2}$}{equal}
\icmlauthor{Maddie Yang$^{4}$}{}
\icmlauthor{Li Li$^{4,5}$}{}
\icmlauthor{Lauren T. May$^{2}$}{}
\icmlauthor{Geoffrey I. Webb$^{1}$}{}\\
\icmlauthor{Shirui Pan$^{3+}$}{}
\icmlauthor{George Church$^{4,5+}$}{}
\\ 
1. Department of Data Science and AI, Monash University
\\
2. Drug Discovery Biology, Monash Institute of Pharmaceutical Sciences, Monash University
\\
3. School of Information and Communication Technology, Griffith University
\\
4. Harvard Medical School, Harvard University
\\
5. Wyss Institute for Biologically Inspired Engineering, Harvard University
\\
* indicates equal contribution and
$+$ indicates corresponding authors:
\\
$\rm George\ Church(george\_{church}@hms.harvard.edu)$, $\rm Shirui \ Pan (s.pan@griffith.edu.au)$


\end{icmlauthorlist}

\icmlkeywords{Machine Learning, ICML}

\vskip 0.3in]




\begin{abstract}
The integration of Large Language Models (LLMs) into the drug discovery and development field marks a significant paradigm shift, offering novel methodologies for understanding disease mechanisms, facilitating drug discovery, and optimizing clinical trial processes. This review highlights the expanding role of LLMs in revolutionizing various stages of the drug development pipeline. We investigate how these advanced computational models can uncover target-disease linkage, interpret complex biomedical data, enhance drug molecule design, predict drug efficacy and safety profiles, and facilitate clinical trial processes. Our paper aims to provide a comprehensive overview for researchers and practitioners in computational biology, pharmacology, and AI4Science by offering insights into the potential transformative impact of LLMs on drug discovery and development. 
\end{abstract}

\section{Introduction}
\label{intro}
\textit{``Language is only the instrument of science, and words are but the signs of ideas.''}

\hfill ------ \textit{Samuel Johnson}

\vspace{5mm}

\begin{figure*}
    \centering
\includegraphics[width=0.95\textwidth]{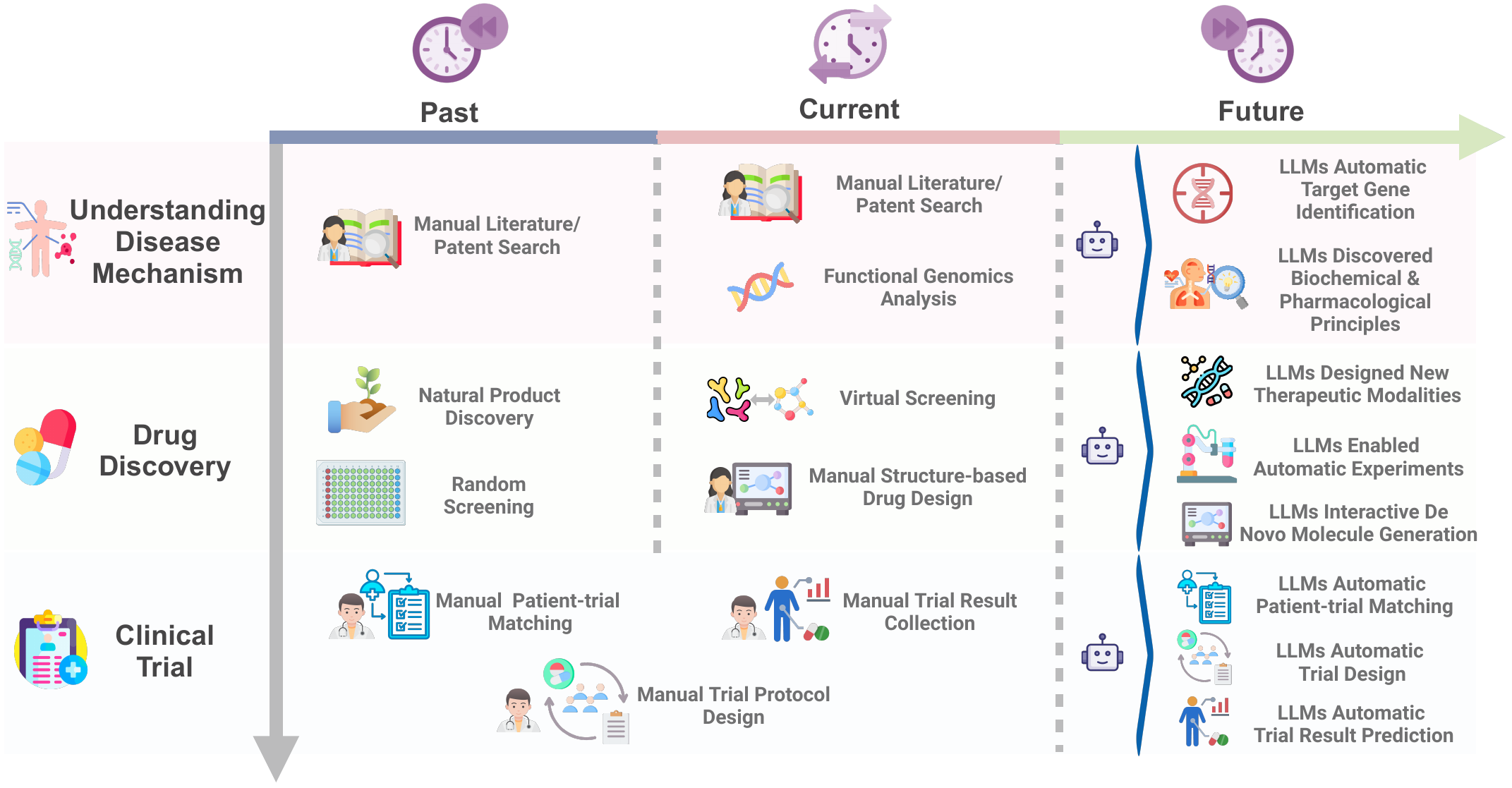}
    \vspace{-6mm}
    \caption{\textbf{Large Language Models Shaping the Future Landscape of Drug Discovery and Development.} In the past, each stage of drug discovery involved numerous manual tasks, which requires significant human effort and substantial resources. Nowadays, advancements in biotechnology, alongside the integration of AI and computer-aided in silico computation tools, have reduced the need of human labor and resources. However, we have yet to have a highly automated drug discovery pipeline, especially in the clinical trial phase, where trail design and matching are still mainly done by clinical practitioners. In the future, it is anticipated that the continued development of LLMs and their application in drug discovery will enable a highly automated drug discovery process. }
    \label{fig:timeline}
\end{figure*}

The pursuit of new drugs to research and develop is a long-term commitment that typically takes 10-15 years and costs over \$2 billion in order to bring a new drug to a patient~\cite{berdigaliyev2020overview}. This complex procedure is traditionally divided into three stages: the first stage is to  understand the disease and to choose the target of treatment; the second stage is to develop a focused approach to developing treatments towards the target; and the third stage is to test the treatments in clinical trials for their effectiveness. Each phase of the process is both time-consuming and resource-intensive, this is because of the complexity of biological systems and the extensive nature of the review required of each phase in the research and validation process. The slow and protracted nature of the process often prevents the introduction of new therapies that would improve and extend human life. Consequently, there are extraordinary dividends to be reaped by introducing efficiencies and expanding the capabilities of current practices.

Artificial intelligence (AI) tools have emerged as preeminent innovation in the quest to accelerate drug discovery and development. Among these tools, large language models (LLMs) \footnote{\noindent Large Language Models (LLMs) are also known as large pre-trained language models.} have distinguished themselves through their capabilities in understanding scientific language and executing various downstream tasks essential in drug discovery and development. Recent LLM breakthroughs,
Geneformer~\cite{theodoris2023transfer}, pretrained on 30 million single-cell transcriptomes, can help in disease modeling and successfully identified candidate therapeutic targets for cardiomyopathy via in silico deletion. Notable LLMs for facilitating chemistry experiments, \citeauthor{boiko2023emergent}~(\citeyear{boiko2023emergent}) and Chemcrow~\cite{bran2023chemcrow}, have highlighted the potential of LLMs in automating chemistry experiments related to drug discovery, specifically in the fields of directed synthesis and chemical reaction prediction. Other works, such as LLM4SD~\cite{zheng2023large}, showed that LLMs can perform scientific synthesis, inference, and explanation directly from raw experimental data and formulate hypotheses that resonate with human experts' analysis. Med-PaLM~\cite{singhal2023large}, a mega size LLM encoding clinical knowledge, was the first to reach human expert in USMLE-styled questions, a medical licensing examination. This advancement highlights the potential of LLMs to liberate clinical practitioners from the laborious activities associated with clinical trials.

With advancements in LLMs, these technologies have the potential to revolutionize the drug discovery pipeline, with future drug discovery including highly automated LLM applications across the three stages of drug discovery (Figure~\ref{fig:timeline}). To understand disease mechanisms and aid in target identification, LLMs can perform comprehensive literature reviews and patent analyses to explore the biological pathways involved in diseases. Additionally, they can conduct functional genomics analysis to pinpoint target genes. By analyzing gene-related literature, including results from in vivo or in vitro experiments, LLMs can compare data on various genes and recommend those with favorable characteristics, such as a desirable mechanism of action or strong potential as drug targets. Furthermore, through analysis and review of literature, LLMs may infer new insights and uncover principles of biochemistry and pharmacology. In the drug discovery and development phase, LLMs have the potential to automate related chemistry experiments by understanding chemical reaction and controlling robotic equipments. In addition, LLMs can offer an interactive platform aiding experts in discovering novel and effective compounds through suggestions for molecule editing and generation. LLMs could also assist in the design of new therapeutic approaches, such as gene therapy. For instance, LLMs could help in the design of Adeno-associated virus (AAV) vectors by quickly summarizing scientific literature to identify novel strategies and by analyzing genomic sequences to predict the most effective vector sequences for safe and efficient gene delivery. During the clinical trial phase, LLMs could streamline the tedious tasks of matching patients with trials and designing trials by interpreting patient profiles and trial requirements. Additionally, early research has shown that LLMs might be capable of predicting trial outcomes by examining historical clinical data.

\textbf{In this survey, we aim to comprehensively address three  questions} for researchers and practitioners seeking to harness the power of LLMs to improve the drug discovery and development pipeline:

\begin{figure*}[ht]
    \centering
    \includegraphics[width=1.0\textwidth]{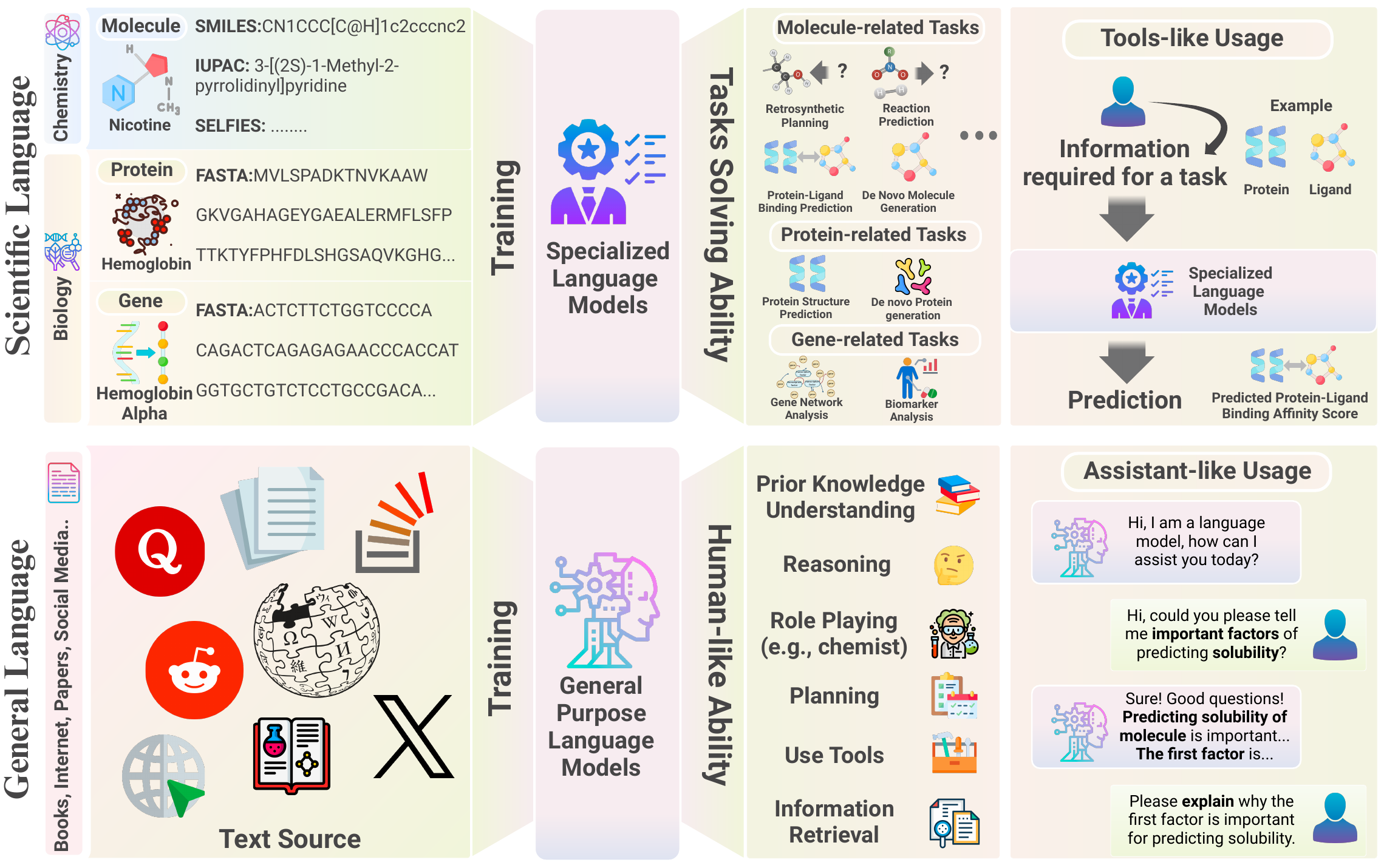}
    \caption{
  \textbf{The two main paradigms of language models.} Specialized language models are trained on specific scientific languages and are typically tailored for specific or a few science-related tasks. These models are used as tools to perform a specific task, in which users provide the information required for a task, and the model outputs the prediction. General-purpose language models are trained on diverse textual information sourced from various materials, including scientific papers and textbooks. These models are used like an assistant that allows users to use plain language to interact with the model.}
    \label{fig:language_models_architecture}
    \end{figure*}

\begin{figure*}
    \includegraphics[width=1.0\textwidth]{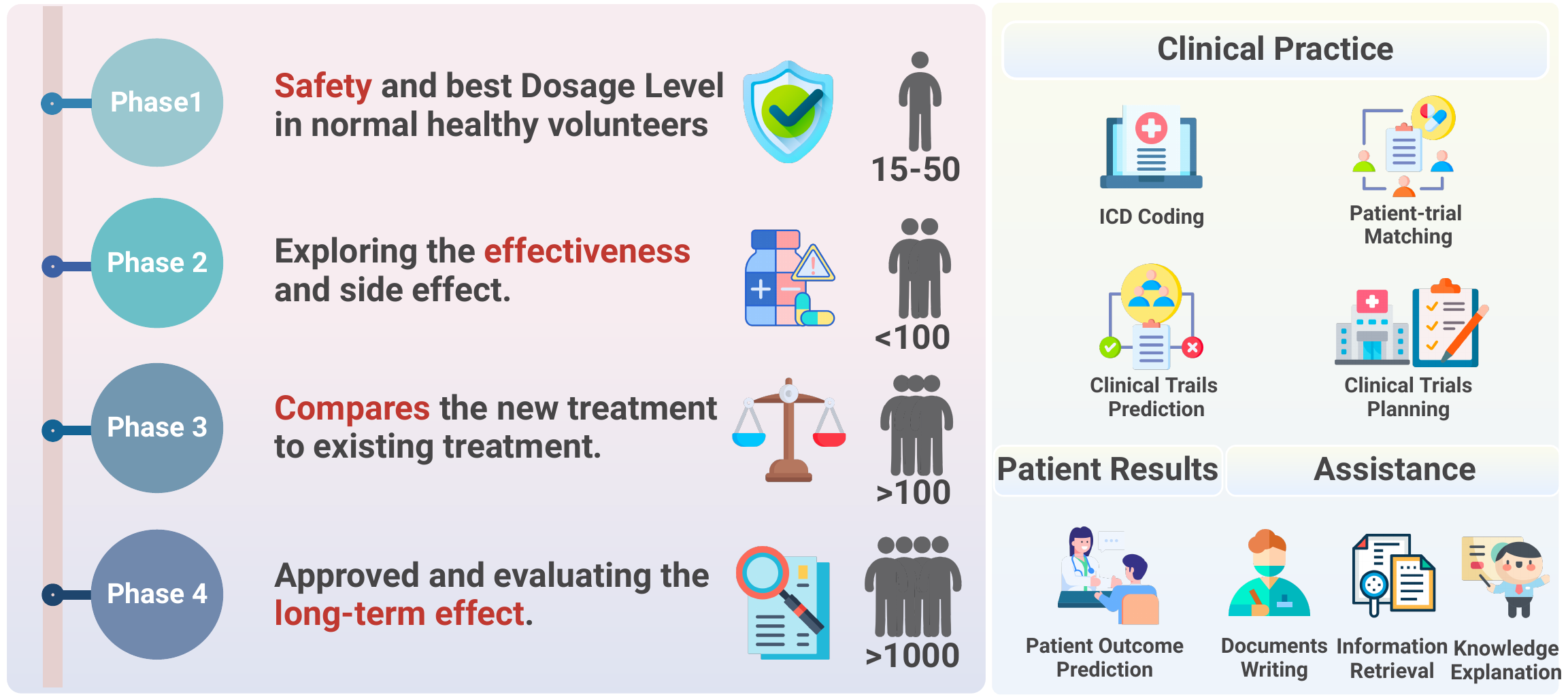}
    \vspace{-8mm}
    \caption{\textbf{Clinical Trials.} The left part of the figure illustrates the processes involved in clinical trials. Clinical trials consist of four phases: Phase 1, Phase 2, Phase 3, and Phase 4. The right part of the figure highlights the tasks that LLMs can perform to facilitate these processes. }
    \label{fig:clinical_trials}
\end{figure*}

\textbf{1) How can LLMs be effectively integrated into various drug discovery and development stages?} First, the types of LLMs under consideration must be defined  (Figure~\ref{fig:language_models_architecture}). Then, we categorize the whole drug discovery and development pipeline into three linear stages: ``Understanding Disease Mechanisms'' (Figure~\ref{fig:understanding_diseases}), ``Drug Discovery'' (Figure~\ref{fig:drug_discovery}), and ``Clinical Trials'' (Figure~\ref{fig:clinical_trials}) to illustrate the blueprint of integrating LLMs into these processes, respectively. The left column of each figure describes the specific processes involved in these stages, while the right column covers the tasks that LLMs can perform to facilitate these stages. The visual representation seeks to illustrate how LLMs can optimize various aspects of drug development.


\textbf{2) How advanced are LLMs in facilitating downstream tasks across various drug discovery and development stages?} In order to determine the level of advancement of LLMs in supporting downstream tasks throughout different stages of drug discovery and development, we have evaluated current applications of LLMs and classified each one into one of four categories: not applicable, nascent, advanced, and mature. These indicators provide an overview of the current state in the field and indicate promising future directions (Figure~\ref{fig:maturity_assessment}).

\textbf{3) What are the future directions of LLMs in drug discovery and development?} We explore the evolving landscape of LLM development, promoting LLMs in more biological use cases, and addressing ethical, privacy, fairness, and bias concerns in future development. These concerns are increasingly apparent as LLMs are applied to handle sensitive health data and make critical medical decisions. We also discuss the need to overcome certain technical limitations associated with LLMs, such as the occurrence of hallucinations, the constraints of context window limit, and the need for better model interpretability and scientific understanding. Solving these challenges can enable LLMs to become trusted and efficient tools in the applications of drug discovery as well as in patient care. This is discussed in Section~\ref{sec:future}.

In this paper, we first elucidate the two paradigms of LLMs for drug discovery and development: specialized language models trained on specific scientific languages and general-purpose language models trained on general textual language. We then delve into how LLMs can be helpful throughout each drug discovery and development stage, from understanding disease mechanisms to facilitating drug discovery and optimizing clinical trials. After covering each stage, we analyze the maturity of these LLM applications. Lastly, we discuss the future directions for LLMs in drug discovery and development.

\section{Main Paradigms of Language Models}
In drug discovery and development, the intricate text-based scientific languages used to describe chemicals and proteins, such as SMILES strings for encoding molecular structures \cite{weininger1988smiles}, and FASTA format for encoding protein, DNA and RNA sequences \cite{fasta}, represent a unique form of structured language crafted by humans to encode domain-specific knowledge. To effectively interpret these languages, two main language model paradigms, including specialized language models (specialized LLMs) and general-purpose language models (general LLMs) emerged (Figure \ref{fig:language_models_architecture}). 

\subsection{Specialised Large Language Models}
The first paradigm of language models in drug discovery and development is specialized LLMs trained in specific scientific languages. These LLMs aim to decode the statistical patterns of scientific language, thereby enabling the interpretation of scientific data in its raw form (Figure~\ref{fig:language_models_architecture}).

\textbf{Understanding Disease Mechanisms.} Specialized LLMs can be used in many ways to explore diseases. For instance, LLMs can extract genomic information from single-cell RNA transcriptomic data and DNA sequences~\cite{consens2023transformers}, enabling practitioners to determine epigenetic marks, transcription factor binding sites, functional genetic variants, and gene network analysis, all of which contribute to understanding the genetic basis of disease.

Additionally, protein LLMs, such as ESM~\cite{rives2021biological}, can be trained to predict parts of the amino acid sequence that have been intentionally hidden or `masked' during training, such as ``MVL$<$MASK$>$PAD” (Figure~\ref{fig:language_models_architecture}). Despite a simple training procedure, these specialized LLMs have been proven helpful in annotating the functions~\cite{matic2022precogx} and predicting the structures of proteins directly from protein sequences~\cite{lin2023evolutionary}, significantly advancing our understanding of protein structures, and informing downstream drug discovery efforts. 

\begin{figure*}
    \includegraphics[width=1.0\textwidth]{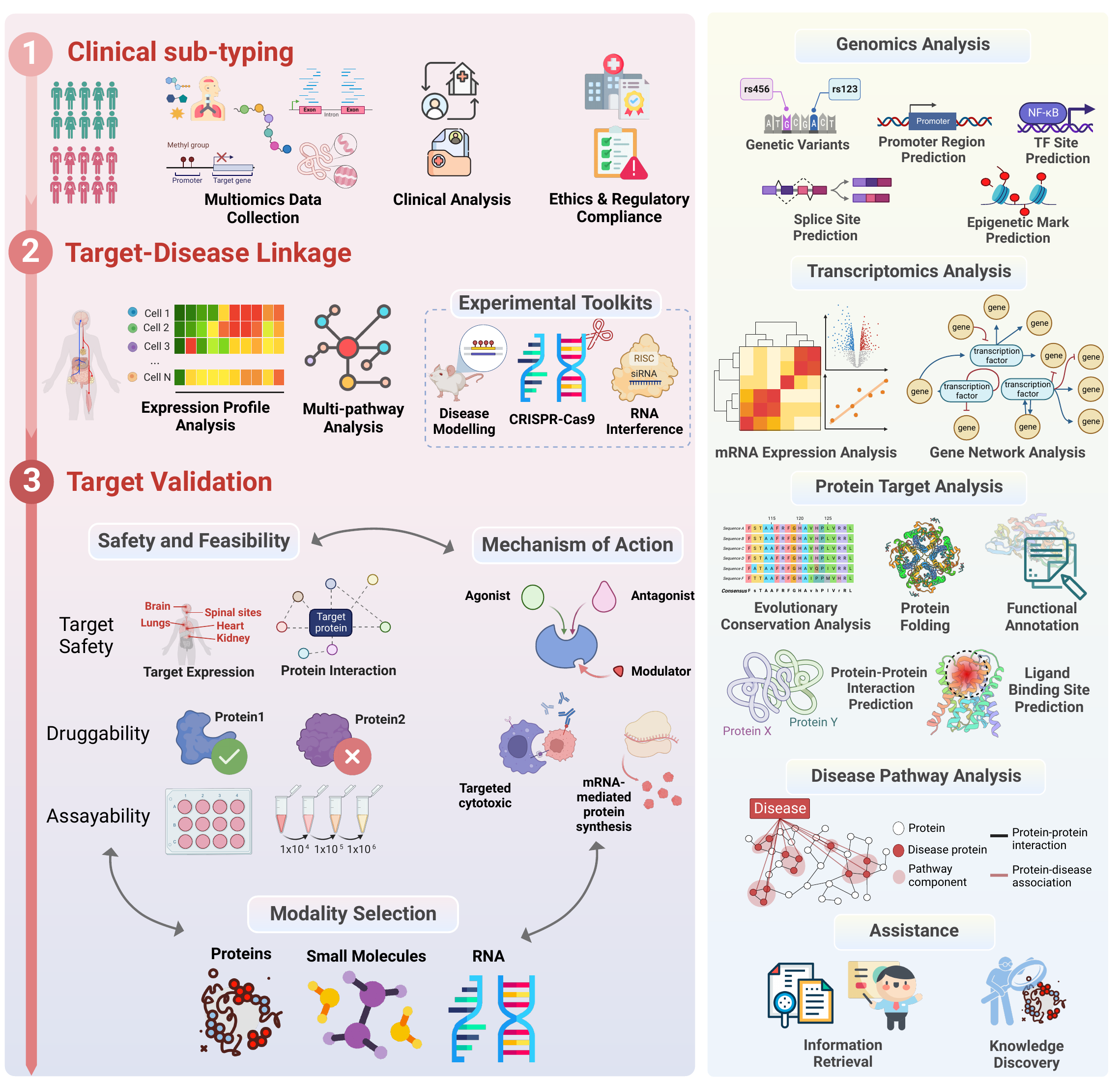}
    \caption{\textbf{Understanding Disease Mechanisms.} The left part of the figure illustrates the processes involved in understanding disease mechanisms. This process involves clinical sub-typing, target-disease linkage analysis, and target validation. Clinical subtyping refers to identifying subgroups of patients with similar clinical characteristics during which data can be collected from multi-omics. Target-disease linkage analysis refers to identifying the relationship between targets and diseases. Target validation typically involves three steps: safety and feasibility, mechanisms of action, and modality selection. The right part of the figure highlights the tasks that LLMs can perform to facilitate these processes, including genomics analysis, RNA analysis, pathway analysis, target profiling, strategic profiling, and assistance.}
    \label{fig:understanding_diseases}
\end{figure*}

\textbf{Drug Discovery.}
In drug discovery, specialized LLMs are particularly helpful for accelerating various chemistry experiments~\cite{bran2023chemcrow,park2023automated}. A specialized LLM trained in the molecular SMILES language can help in retrosynthetic planning and predicting reaction outcomes; at the same time, it can help chemists in de novo molecules guided by some specific molecular properties, such as increasing binding affinity towards targets. Moreover, these models can also play a role in ADMET (Absorption, Distribution, Metabolism, Excretion, and Toxicity) prediction, a critical step in assessing molecule properties and filtering out those with undesirable characteristics.


\textbf{Usage.}
Specialized LLMs are tool-like, where the user inputs information needed for a given task and receives a model prediction in return (Figure~\ref{fig:language_models_architecture}). For instance, when we use a specialized LLM to query protein-ligand binding affinity, both protein sequences and ligand SMILES strings must be provided to the model, which will subsequently output the predicted binding affinity score.

\subsection{General-purpose Language Models}
The second paradigm encompasses general LLMs, which are trained on a diverse array of textual information sourced from various materials, including but not limited to scientific papers, textbooks, and general literature. Such breadth in training allows them to achieve a broad understanding of human language, which includes a significant grasp of scientific contexts. Models like GPT-4~\cite{gpt4-technical-report, ai4science2023impact} and Galactica~\cite{taylor2022galactica} have been noted for their proficiency in also mastering complex formal scientific description languages, including SMILES strings and FASTA format. Using this capacity, general LLMs can work on tasks that would typically require the participation of domain professionals, such as making inferences, doing reasoning and analysis, and applying field-specific knowledge across different scientific domains.

\textbf{Understanding Disease Mechanisms.}
General LLMs can traverse a large volume of literature, extract data, and summarize for users. Furthermore, it can also synthesize the extracted data into a knowledge graph, revealing how genes and diseases are connected, helping scientists uncover the basis behind diseases~\cite{savage2023drug}. Furthermore, these models can explain technical terminologies in layperson's language, making understanding complex concepts and principles easy, significantly aiding in education and communication.

\textbf{Drug Discovery.}
In drug discovery, general LLMs have great potential to accelerate experimental practices. Recently, general LLMs have been applied in chemistry robotics for automated experiments. General LLMs also exhibited expert-level capabilities in retrosynthetic planning and reaction prediction~\cite{bran2023chemcrow, boiko2023emergent}, while only costing a fraction of human experts. 

Some preliminary attempts are underway to train general language models to perfect tasks currently more suited for specialized LLMs, such as de novo molecule and protein generation and editing~\cite{liu2023chatgpt}. The primary motivation is that, unlike specialized LLMs that can only learn data patterns from specific scientific languages, these LLMs can reason and apply the domain knowledge learned from extensive literature. However, research in this direction is still in its infancy. 

\textbf{Clinical Trials.}
General LLMs provide significant advantages in analyzing electronic health records and clinical protocols~\cite{singhal2023large,jin2023matching,huang2019clinicalbert}. They can facilitate patient-trial matching, assist in trial planning, help predict trial outcomes, and assist in document writing. The user-friendly chat interfaces of general LLMs also make it easier for practitioners to interact with them.

\textbf{Usage.}
A natural language-based AI Assistant is a strong use case for general LLMs (Figure~\ref{fig:language_models_architecture}). For example, if a user wanted to know what some key features are in predicting the solubility of molecules, they could ask a general LLM, and it would retrieve and summarize relevant information from the literature.

\section{LLMs in Drug Discovery and Development}
This section discusses how LLMs can be applied in three drug discovery and development pipeline stages: understanding disease mechanisms, drug discovery, and clinical trials. 

\begin{figure*}
    \includegraphics[width=1.0\textwidth]
    {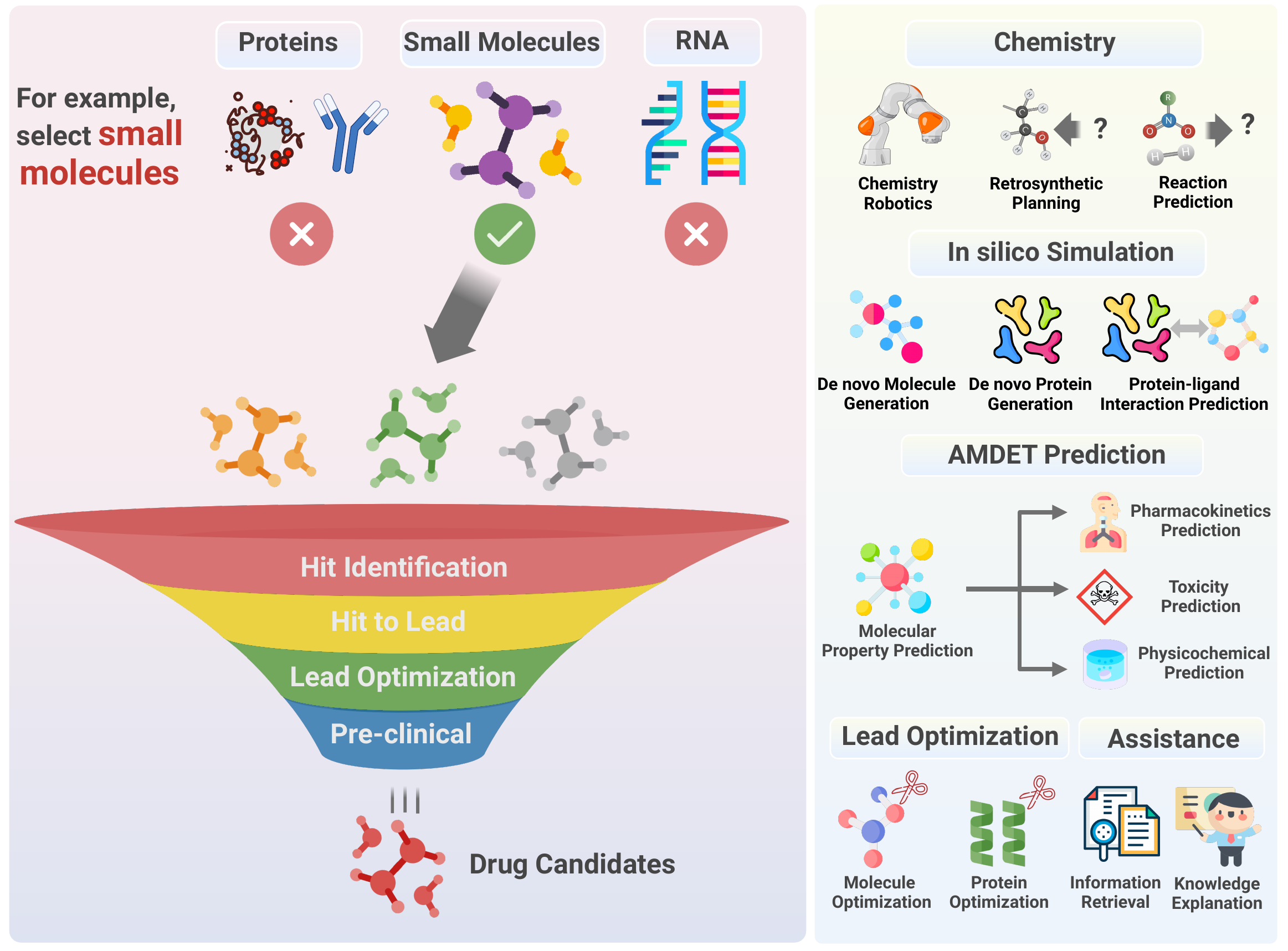}
    \vspace{-8mm}
    \caption{\textbf{Drug Discovery.} The left part of the figure illustrates the processes involved in drug discovery. The right part of the figure highlights the tasks that LLMs can perform to facilitate these processes. }
    \label{fig:drug_discovery}
    \vspace{-2mm}
\end{figure*}

\subsection{Understanding Disease Mechanisms}
Understanding disease mechanisms is the initial and crucial stage of the drug discovery and development pipeline. The primary aim of this stage (Figure~\ref{fig:understanding_diseases}) is to identify a suitable protein target for a potential drug to act upon~\cite{lindsay2003target}. This process involves three key steps: clinical sub-typing, target-disease linkage analysis, and target validation. 

Clinical sub-typing in drug discovery involves categorizing patients into subgroups to collect clinical and multiomics data and aids in understanding disease variations and identifying potential differences in disease mechanisms across patient groups~\cite{cortes2013path,pun2023ai}. 

The target-disease linkage analysis phase in drug discovery involves establishing connections between potential protein targets and specific diseases. This phase encompasses pathway analysis to investigate biological pathways involved in the disease and expression profile analysis to study disease-related gene expression patterns~\cite{plenge2013validating}. Additionally, practitioners leverage experimental techniques to establish causal links between a target and the disease, including CRISPR-Cas9~\cite{lin2017crispr}, in-vivo disease modeling~\cite{lindsay2003target}, and interference RNA (siRNA)~\cite{cortes2013path}.

After identifying a target in the drug discovery process, target validation is a crucial, non-linear step that follows target identification, involving a continuous validation cycle with no fixed starting point (Figure~\ref{fig:understanding_diseases}). This cycle includes assessing the necessary actions to be performed on the target for disease treatment (mechanism of action), choosing the most appropriate therapeutic intervention (modality selection), and conducting a comprehensive safety and feasibility assessment~\cite{emmerich2021improving}. The safety and feasibility assessment evaluates both the potential organismal impact (safety) and the target's druggability~\cite{floris2018genetic}, as well as the practicality of assays for feasibility~\cite{vincent2015developing}. This flexible, iterative approach ensures thorough evaluation of the target's viability and safety at any stage before advancing in drug development, ensuring the selected targets are both theoretically promising and practical for further development.

\subsubsection{Genomics Analysis}
Decades of genome-wide association studies (GWAS) have identified critical genomic regions linked to various diseases~\cite{michailidou2015genome,michailidou2017association,nelson2017association,zengini2018genome} that have significantly advanced genomic-based analysis for disease understanding and target discovery. Notably, integrating genetic associations in drug discovery, could significantly improve the success rate of clinical targets~\cite{nelson2015support}. 

Recently, there has been significant interest in adapting advancement in LLMs used for human languages to genomic analysis, such as DNA-BERT~\cite{ji2021dnabert}, due to the structural similarities between DNA and human language. Through specialized training on vast amounts of nucleotide sequences, these LLMs are adept at decoding the language of genetics. As a result, there has been an explosion in the field of specialized nucleotide LLMs~\cite{ji2021dnabert} that are increasingly capable of understanding the cryptic ``language'' used by genomes more efficiently, enabling various downstream tasks in understanding genetic mechanisms of diseases. 

\textbf{Genetic variant analysis.}
The application of nucleotide LLMs in genetic variant analysis hinged on the fact that genetic sequences follow specific language patterns and rules~\cite{yanofsky1964protein,altschuh1988coordinated}. Variations in these sequences—be it single nucleotide polymorphisms (SNPs), insertions, deletions, or more complex rearrangements—can significantly impact gene function~\cite{brendel1984genome,searls2002language}. Hence, they employ masked language modeling when nucleotide LLMs are trained on extensive genomic data. In this approach, the model learns to predict parts of the nucleotide sequence that have been intentionally hidden or `masked' during training. This learning process enables the LLMs to decode the intricate, often hidden patterns and rules that govern the language of genes~\cite{ji2021dnabert,dalla2023nucleotide}.

Post-training, nucleotide LLMs have demonstrated the ability to detect significant functional genetic variants directly from DNA sequences. For example, DNA-BERT~\cite{ji2021dnabert} showed that nucleotide LLMs selectively concentrate on the most relevant genomic regions. This enables the extraction of motif patterns that are evolutionarily conserved and aid in identifying functional variants of significance. Similarly, Nucleotide Transformer~\cite{dalla2023nucleotide}, another specialized nucleotide LLM, has also demonstrated the ability to prioritize functional genetic variants. Moreover, it can be further trained for specific variant identification tasks, such as classifying SARS-CoV-2 variants~\cite{zhou2023dnabert} and understanding SARS-CoV-2 evolutionary dynamics~\cite{zvyagin2023genslms}.

More recently, HyenaDNA, built on the Hyena LLM framework, has pushed the boundaries of genetic variant analysis by enabling the modeling of extremely long genomic sequences—up to 1 million tokens—at the single nucleotide level~\cite{nguyen2024hyenadna}. This is a significant leap from previous models that were constrained by the quadratic scaling of attention and limited to much shorter sequences. HyenaDNA’s ability to process such extensive context lengths allows it to capture long-range interactions in DNA, which are crucial for understanding complex genetic variations. It has achieved state-of-the-art performance on multiple benchmarks with a much smaller model and less pretraining data, marking a substantial advance in the field of genomic sequence analysis. This long-range capability, coupled with the precision of single nucleotide resolution, positions HyenaDNA as a powerful tool in detecting and prioritizing functional genetic variants, further enhancing our understanding of genomic data.

\textbf{Genomic regions-of-interest predictions.}
Promoter regions, transcription factor (TF) binding sites, and splice sites are all crucial elements in regulating gene expression. Despite their varied roles, they all contribute to the complex regulation of when, where, and how genes are activated or silenced. Alterations or mutations in these regions may bring about overexpression or underexpression of a gene, potentially causing diseases. However, despite the importance, predicting these regions remains challenging due to the need for more understanding in the language of DNA~\cite{ji2021dnabert}.  

\begin{figure}
    \centering
    \vspace{-3mm}
    \includegraphics[width=0.5\textwidth]{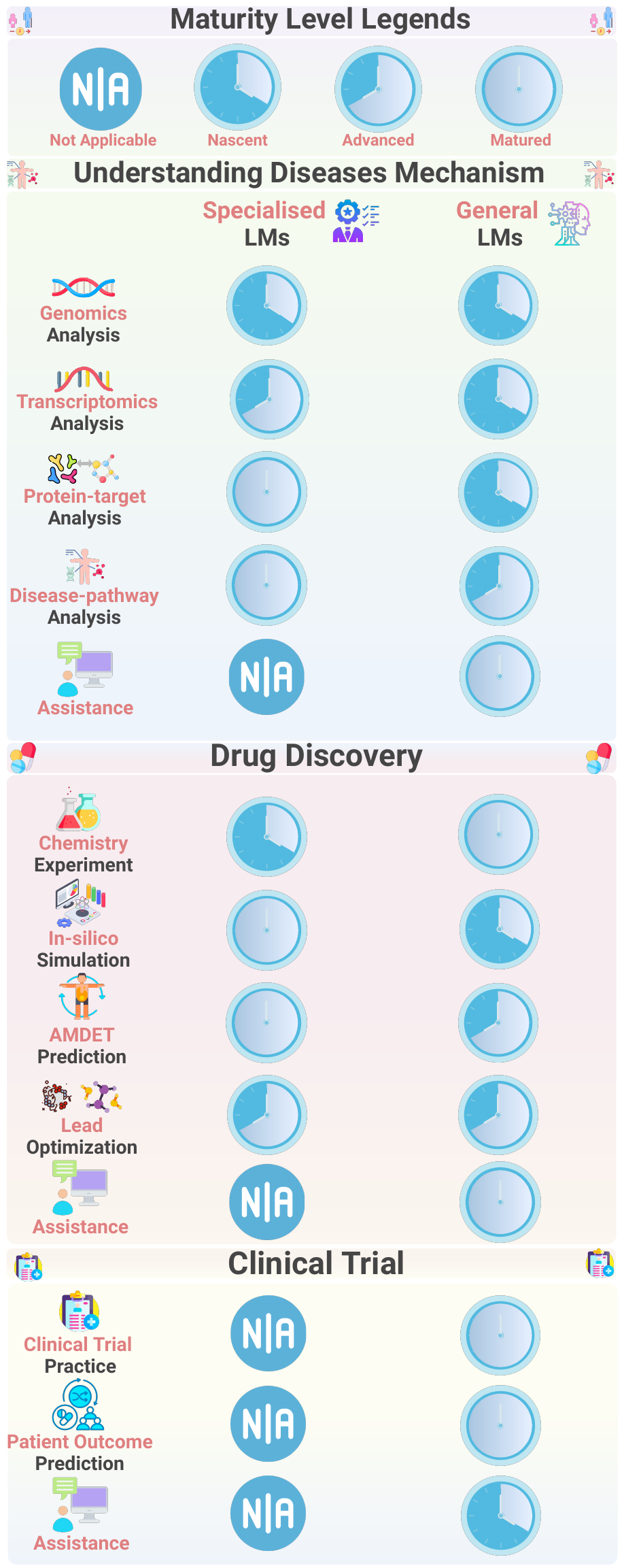}
    \vspace{-9mm}
    \caption{\textbf{Maturity Assessment of LLMs in Downstream Tasks.} This figure is segmented into Maturity Level Legends, Understanding Disease Mechanism, Drug Discovery, and Clinical Trials, detailing LLMs maturity across various tasks and phases.}
    \label{fig:maturity_assessment}
\end{figure}

To address these challenges, specialized nucleotide LLMs are being fine-tuned to predict these regions of interest, with results showing an outperformance against previous state-of-the-art methods~\cite{dalla2023nucleotide, zhou2023dnabert}. Fine-tuning these LLMs for genomic applications involves two key steps. First, an LLM is pre-trained on extensive datasets of nucleotide sequences, allowing the LLM to grasp the nuances of genetic language. Then, in the second stage, the LLM is further fine-tuned to incorporate domain-specific knowledge, which is the locations, biological functions, and additional biochemical and biophysical insights of different gene regulatory sites. 

Epigenetic marks refer to chemical modifications, such as DNA methylation and histone modifications, that affect gene expression without altering the underlying DNA sequence. These marks play a crucial role in regulating gene activity, influencing both disease development and therapeutic targeting~\cite{atlasi2017interplay}. Accurately predicting these marks is essential for understanding how epigenetic changes impact gene expression and their implications in various diseases~\cite{miranda2019epidrugs}. However, the complexity and variability of epigenetic marks present a significant challenge in making accurate predictions. Similar to the discussion above, pre-trained nucleotide LLMs are further fine-tuned to predict specific histone modifications, such as H3K14ac, H3K36me3, and H3K4me1~\cite{zhou2023dnabert}.

\subsubsection{Transcriptomics Analysis.}
Transcriptomics, a field that investigates the entirety of RNA transcripts that an organism or cell system generates under certain conditions, has experienced a surge in the volume of transcriptomic data derived from a wide range of human tissues due to the development of high-throughput technologies and single-cell technologies. However, the data is often sparse for specific disease states, particularly for rare diseases and diseases affecting clinically inaccessible tissues~\cite{shao2021scdeepsort}, so relying solely on these data for specific diseases would likely not suffice to develop robust and accurate models. To address existing limitations, specialized gene LLMs have been proposed to obtain a comprehensive understanding of transcriptomic data while offering the goods to adapt to scenarios with sparse data samples.

The primary technological development in this subfield is the specialized transcriptomic LLM, Geneformer~\cite{theodoris2023transfer}, which developed an innovative method for mapping each single-cell transcriptome into a sequence of genes ranked by their expression levels. This approach, known as ``rank value encoding'', represents the transcriptome of each cell as a sequence of genes ordered based on their expression levels. These levels are then normalized against the overall expression observed across all human tissues. Through this technique, rank value encoding offers a distinct representation of gene activity within individual cells and facilitates a comprehensive comparison of gene expression across a diverse array of data~\cite{theodoris2023transfer}. This approach is akin to learning the language of transcriptomics through specialized LLMs, enhancing our understanding of cellular behaviors and interactions at the molecular level. 

By transforming single-cell transcriptomic data into gene sequences, Geneformer~\cite{theodoris2023transfer}, scGPT~\cite{cui2023scgpt}, and other models like scMulan~\cite{bian2024scmulan} and scFoundation~\cite{hao2024large} have demonstrated a remarkable ability to analyze transcriptomic data effectively as foundational LLM models. Furthermore, specialized transcriptomic LLMs can be adapted to scenarios with sparse data through fine-tuning to model gene networks accurately to comprehend complex dynamics, including network interactions, extending beyond simple cell-level annotations~\cite{ma2024harnessing}. 

In parallel, efforts have also been made to leverage biomedical literature for predicting future therapeutic targets by training LLMs on historical text corpora. A study used Word2Vec models on abstracts published between 1995 and 2022, allowing these LLMs to prioritize gene-disease associations and protein-protein interactions likely to be validated in future research \cite{narganes2023publication}. This approach, termed Publication-Wide Association Study (PWAS), encodes biomedical knowledge as word embeddings without human supervision, effectively capturing drug discovery concepts and prioritizing hypotheses years before experimental confirmation. PWAS demonstrates the potential of LLMs as a scalable system for early-stage target ranking, enhancing the ability to mine literature for under-explored therapeutic opportunities.

\textbf{mRNA expression analysis.}
mRNA expression analysis can be challenging due to the need to derive meaningful insights from limited data scenarios, a vital aspect in enhancing our understanding of diseases. Traditional machine learning approaches, such as XGBoost~\cite{chen2016xgboost} and standard deep neural networks, usually start from scratch for each specific task. This methodology can be ineffective, especially if the data is limited, which is often the case in the fields of rare disease research or when working with tissues that are not accessible in a clinical setting.

The specialized transcriptomic LLM, Geneformer~\cite{theodoris2023transfer}, leverages the general knowledge acquired from pretraining on transcriptomic data to adapt efficiently to a specific disease use case and demonstrates remarkable efficiency in gene network analysis with minimal data. A study successfully distinguished key factors in the NOTCH1-dependent network by fine-tuning on just 884 endothelial cells from healthy versus dilated aortas, outperforming other methods that used a much larger dataset of about 30,000 cells~\cite{theodoris2023transfer}. 

scGPT~\cite{cui2023scgpt}, another generated pre-trained transformer model for single-cell multi-omics data analysis, on the other hand, also showed the ability to generate meaningful cell-type clusters directly from the pre-trained model in a zero-shot manner (i.e., without additional fine-tuning). 

\textbf{Gene network analysis.}
Gene network analysis typically begins by mapping the gene regulatory networks and tracing the critical genes in a disease's progression to identify potential therapeutic targets. Specifically, the gene network analysis strives to uncover vital regulatory elements that can alter or modulate these networks in a desired manner~\cite{theodoris2015human,theodoris2021network}. Due to the lack of data, these regulatory elements are challenging to find, especially for rare diseases or conditions affecting clinically inaccessible tissues~\cite{cui2023scgpt,theodoris2023transfer}. 

To address the challenge of gene network analysis, Geneformer~\cite{theodoris2023transfer} proposed leveraging the self-attention mechanism within the transformer~\cite{vaswani2017attention} backbone of the LLM to address the challenge of gene network analysis. This self-attention is crucial, as the trained attention weights of the model for each gene reveal two essential aspects: (1) the genes to which a particular gene is paying attention and (2) the genes that focus on it. This process inherently constructs a gene network, intricately mapping the web of gene interactions. Geneformer~\cite{theodoris2023transfer} also employs ``in silico deletion'', a technique similar to perturbation analysis for virtually removing specific genes to study their impact on gene networks. scGPT~\cite{cui2023scgpt}, on the other hand, uses embedding computing to construct a gene network via gene-gene similarities. 

\subsubsection{Protein Target Analysis}
A protein sequence is often the most accessible data about a target that can help explain its potential to play a role in disease mechanisms, with drug discovery scientists often targeting it as a starting point. In this application, specialized LLMs can be particularly valuable as these models have shown the ability to provide extensive analyses, including evolutionary conservation, functional annotation, protein folding, and binding site prediction. The unique ability of the LLMs to extract relevant information from sequence data alone has provided a means of characterization of target biological traits and functions even without experimental data like experimentally determined 3D protein structures.

\textbf{Evolutionary conservation.}
The use of specialized LLMs in protein analysis, as explained in the representative work of ESM \cite{rives2021biological}, is based on a fundamental idea: the statistical patterns of protein sequences contain valuable information about their biological function and structure, which have been shaped by evolutionary processes~\cite{yanofsky1964protein, altschuh1988coordinated}. This idea proposes that mutations that improve an organism's fitness are more likely to be selected by evolutionary forces among the multitude of possible mutations a sequence can undergo~\cite{gobel1994correlated}, resulting in unique signatures in protein sequence patterns.

Specialized protein-based LLMs can effectively predict likely mutations within protein sequences that contain masked amino acids. This proficiency not only enables them to understand evolutionary conservation but also allows them to grasp the selection processes driving the evolution of these sequences. Follow-up research has demonstrated the efficacy of this approach using the ESM language model, which can make accurate predictions of mutational effects across a variety of proteins with different functions without any additional training~\cite{meier2021language}. MSA-Transformer takes this approach further by analyzing multiple sequences using a multiple sequence alignment (MSA) approach~\cite{rao2021msa}. With the added information from the MSA as input, MSA-Transformers enhance their ability to interpret complex relationships within protein sequences and outperform single-sequence LLMs~\cite{meier2021language}.

Understanding evolutionary conservation using specialized LLMs is significant because it provides information not only on the functional landscape of proteins according to amino acid conservation patterns but also enables the establishment of the role and significance of individual residues in binding and activity~\cite{altschuh1987correlation}. This is important as such conserved sites typically contribute to protein function and structure, while covarying mutations are similarly associated with these features, including contact surface, structure, and binding~\cite{levitt1978conformational, yanofsky1964protein, altschuh1988coordinated}. Such findings are crucial for developing specialized LLMs for proteomics and form the basis for using these models to provide insight into predicting protein folding, binding sites, and functional annotation.

\textbf{Protein folding.}
⁤The sequence patterns of a protein are shaped by its hidden structure, which is linked to evolutionary conservation and mutation. ⁤⁤Specific structures and sequences are conserved due to their functional importance, while mutations occur in response to evolutionary pressures. ⁤⁤As a result, LLMs that learn from protein sequence data can indirectly capture these evolutionary trends. ⁤⁤This is exemplified by the groundbreaking work in ESM~\cite{rives2021biological} and MSA-Transformer~\cite{rao2021msa}, which showed that LLMs can accurately decode the structural nuances of proteins from sequence data alone. ⁤⁤Specifically, when these LLMs create pairwise interaction maps (attention matrices) between all amino acid positions in a sequence, they demonstrate an ability to infer which pairs of amino acids should be in contact with unparalleled accuracy~\cite{rao2020transformer,fung2022impact}. ⁤⁤This remarkable ability strongly suggests that a significant amount of structural information can be directly inferred from the LLM model using only sequence data, in line with Anfinsen's dogma~\cite{anfinsen1973principles}. ⁤

⁤Building on foundational research, AlphaFold2~\cite{jumper2021highly} and RosettaFold~\cite{baek2021accurate} have revolutionized the field of protein structure prediction. ⁤⁤These models can now produce atom-level accuracy even in cases where similar structures are not known, due to the Evoformer component in AlphaFold2, a specialized protein-based LLM. ⁤⁤In this way, AlphaFold2's training objective mirrors that of MSA-Transformer, where residues from MSA undergo random masking before being reconstructed by the Evoformer~\cite{jumper2021highly, hu2022exploring}. ⁤⁤Through this process, AlphaFold2 can reason over MSAs and incorporate valuable evolutionary information to achieve near-experimental accuracy at the whole protein's structure level \cite{mirdita2022colabfold, ahdritz2024openfold}. Furthermore, this same approach has been further extended to understanding biomolecular interaction with RosettaFold All-Atom, enabling the modeling of complex biomolecular assemblies, such as protein-protein complexes, protein-DNA/RNA interactions, and protein-small molecule interactions \cite{krishna2024generalized}.

⁤In a similar timeframe, RGN2~\cite{chowdhury2022single} developed ProtBERT to encode protein sequence data and predict structure directly from a single sequence, without requiring evolutionary information from an MSA. ⁤⁤RGN2 has demonstrated the ability to match or even surpass AlphaFold2~\cite{jumper2021highly} in predicting the structure of orphan proteins that lack sequence homologs \cite{chowdhury2022single}. ⁤

\textbf{Functional annotation.}
As elucidated by the early works \cite{rao2019evaluating,rives2021biological,brandes2022proteinbert,lin2023evolutionary}, specialized protein-based LLMs can encode rich structural and functional information~\cite{bepler2021learning}. Naturally, the rich information encoded within LLMs is now being harnessed in advanced applications like NetGO 3.0~\cite{wang2023netgo} to advance automated function prediction of proteins without costly experiments. The same approach was used by \citet{matic2022precogx} for GPCR sequence analysis using ESM~\cite{rives2021biological}, aiming to predict their signaling and functional repertoire. The research showed that the interaction mechanism between different protein variants is due to alternative splicing of genes, and GPCRs can be clearly defined with the help of specialized protein-based LLMs.

With advancements in general LLMs, ProteinChat~\cite{guo2023proteinchat} was proposed to provide an interactive platform where users can upload protein sequences and structures and pose questions about a brief description of a protein's functionalities. A recent work~\cite{ai4science2023impact} also demonstrates that GPT-4 demonstrates considerable expertise in understanding proteins.

Despite these advances, the study of protein language models has remained relatively limited in scope. However, ESM2~\cite{lin2023evolutionary} and ESM3~\cite{hayes2024simulating} have been addressing this gap by scaling up to an impressive 15 billion and 98 billion parameters respectively, making them some of the largest protein language models evaluated to date. This substantial increase in scale has enabled ESM models to better learn the sequence-structure-function relationships of proteins. Using ESM as the foundational LLM, ESMFold has demonstrated an unprecedented ability to closely match the performance of AlphaFold2 in predicting protein structures~\cite{lin2023evolutionary}. Notably, ESMFold achieves this high level of precision by analyzing a single sequence, and its ability to operate without querying a database significantly enhances its speed and ease of use.

\textbf{Protein-ligand interaction and binding site prediction.}
Protein-ligand interactions and understanding protein binding sites play a pivotal role in understanding protein function and interactions and serves as an essential foundation for rational-based drug discovery. Gaining insight into these interactions is crucial in identifying potential safety issues related to target proteins and acts as guidance for the design of therapies. Protein-based specialized LLMs have demonstrated notable successes leveraging their extensive understanding of protein language. These include the prediction of metal ion binding sites~\cite{yuan2022alignment}, protein-protein binding sites~\cite{fang2023deepprosite}, and small molecule binding sites~\cite{zhang2023protein}. Understanding the interactions between proteins is thus fundamentally important in deciding a valid target for treating diseases. Furthermore, this knowledge can contribute to the design of biologics-based drugs, as discussed in sections 3.2.2 and 3.2.4.

To this end, protein-based LLMs have advanced with the introduction of AlphaFold-Multimer~\cite{evans2021protein}. While its initial application is in predicting multimeric complex structures, it turns out that AlphaFold-Multimer is capable of predicting protein-protein interaction directly from protein sequences as accurately as unique protein-protein docking methods that use experimentally-determined structures~\cite{ketata2023diffdockpp}.

Another significant development in protein-based LLMs for protein-protein interaction is DockGPT~\cite{mcpartlon2023deep}, an innovative approach in protein docking. This end-to-end deep learning method stands out for its flexible and site-specific protein docking capability, effectively accommodating conformational flexibility and utilizing binding site information compared to AlphaFold-Multimer. Its strength lies in its ability to process unbound and predicted monomer structures. Notably, DockGPT~\cite{mcpartlon2023deep} showed that the protein-based LLM  can effectively deal with antibody-antigen complexes, achieving high accuracy in predicting binding poses as well as co-design the sequence and structure of antibody regions targeting specific epitopes. 

When it comes to assessing ligand interaction sites, Phosformer has proven to be a significant improvement. ⁤⁤Unlike previous methods such as MusiteDeep, DeepPhos, and Ember~\cite{wang2017musitedeep,luo2019deepphos, kirchoff2022ember}, which relied on multiple models specific to different protein families or groups, Phosformer uses its comprehensive understanding of protein language to make accurate predictions with just one model. ⁤⁤This means that virtually any kinase interacting with any peptide can be used to predict phosphorylation sites. ⁤⁤Meanwhile, ProtT5~\cite{elnaggar2021prottrans}, a specialized protein-based machine learning model, has been used to successfully predict binding sites for metal ions, nucleic acids, and small molecules~\cite{littmann2021protein}. ⁤⁤This approach has even been advanced to predict genome-wide annotations for these binding sites~\cite{yuan2023genome}. ⁤ 

Finally, recent advancements in protein-ligand interaction modeling using LLMs have been significantly enhanced by RosettaFold All-Atom~\cite{krishna2024generalized}. Unlike earlier tools that focused primarily on polypeptide chains, RosettaFold All-Atom incorporates a wide range of ligands, including small molecules, metal ions, and nucleic acids, into its predictions. This comprehensive approach not only enables highly accurate modeling of protein-ligand complexes, but provide analysis of key cellular processes for protein target analysis, offering deep insights into disease mechanisms and providing a powerful tool for drug discovery~\cite{krishna2024generalized}.

\subsubsection{Pathway Analysis}
In pathway analysis, gene regulatory network analysis can be a powerful tool for researchers seeking to descipher complex disease pathways. In this subsection, we will be focusing on the use of general LLMs, which can provide all-around assistance for pathway analysis. 

Unlike their specialized counterparts, general LLMs are innately equipped with a wealth of prior knowledge gleaned from vast scientific literature and datasets~\cite{taylor2022galactica,gpt4-technical-report}. This extensive background enables them to approach pathway analysis with a broad, informed perspective rather than specializing in a single scientific language. Furthermore, general-purpose LLMs have the distinct advantage of being able to interactively and conversationally engage with complex scientific data~\cite{gpt4-technical-report,jeblick2023chatgpt}, providing researchers with a powerful tool for understanding and exploring their findings. 

A recent study showcased the capacity of general-purpose LLMs, such as GPT-4, in analyzing blood transcriptional modules related to erythroid cells, demonstrating that these models are efficient in knowledge-driven pathway analysis~\cite{toufiq2023harnessing}. This research uses general LLMs to automatically generate codes for gene networks, summarize candidate genes ranked based on association tests, generate reports for users, and fact-check the report against the literature. In each task, the rich prior knowledge and interactive capabilities of general LLMs are exploited to analyze scientific data. By leveraging the rich prior knowledge and interactive capabilities of general LLMs, this study highlights how they can enhance disease mechanisms and target identification, allowing for a better understanding of complex gene networks. Ultimately, this transforms pathway analysis from static approaches to more dynamic and interpretable methods.

\subsubsection{Assistance}
⁤⁤The exploration of disease mechanisms is a complex task, requiring the contribution of experts from fields across health and pharmaceutical industries. ⁤⁤In this environment, general-purpose LLMs that can perform tasks related to interactive and conversational skills, including information retrieval and knowledge explanation, can play a crucial role~\cite{taylor2022galactica}. ⁤

⁤General-purpose LLMs offer fast and accurate information retrieval, clear explanations tailored to user needs, and the ability to organize and categorize large datasets, enhancing workflow and productivity~\cite{jeblick2023chatgpt}. By integrating with search engines, recent LLM iterations provide real-time access to scientific data, improving hypothesis generation and validation in disease research. Additionally, they aid in scientific communication by simplifying complex ideas for laypeople, fostering better collaboration among specialists with different expertise.

\subsection{Drug Discovery}
The drug discovery process is a crucial phase in the drug development pipeline, encompassing several critical steps as depicted in Figure~\ref{fig:drug_discovery}. These steps include hit identification, hit to lead, lead optimization, and preclinical development.

The process starts with ``hit identification'', where professionals find compounds with potential therapeutic effects. Next, ``hit to lead'' involves a more refined selection from these hits, identifying those most promising for further development. The third step, ``lead optimization'', is a critical process of enhancing a lead compound's efficacy, stability, and safety via editing. Finally, ``preclinical development'' entails rigorous testing of the optimized lead compound in animal models to assess its suitability for human trials. 

Our survey will begin by outlining the specific downstream tasks associated with each operation. Following this, we will explore how LLMs can be incorporated into these tasks to advance the drug discovery process.

\subsubsection{Chemistry}
Medicinal chemistry is essential to drug discovery and development through independent laboratory work and compound synthesis. Autonomous lab operations use robotic manipulators, controlled and programmed to execute complex chemistry and synthetic reactions. In addition, high-throughput screening requires compounds to be precisely and efficiently synthesized as part of the hit identification phase. After synthesis, the compound will be evaluated for activity and selectivity using pharmacological assays.

LLMs have proven to be highly valuable in these fields. LLms can help generate codes that program the chemistry robotics based on user requirements~\cite{bran2023chemcrow,boiko2023emergent}. In particular, LLMs can translate user requirements into complex experimental protocols and convert them into specific, understandable robot instructions. Additionally, LLMs are successful in retrosynthetic planning and reaction prediction, offering to recommend feasible synthetic routes and to predict possible chemical reactions. Using LLMs in this manner is beneficial as it brings efficiencies in compound synthesis and accelerates the drug discovery process. 

\textbf{Chemistry Robotics.}
Nowadays, chemistry robotics is an integral part of conducting chemistry experiments using autonomous laboratory operations. This technique involves converting instructions written in natural language into robot-executable plans, usually described using a fixed, well-defined language that resembles coding.

General LLMs like GPT-4~\cite{gpt4-technical-report} and CodeLlama~\cite{roziere2023code} have shown the ability to generate effective code. Therefore, it is logical to use  general LLMs to generate robot-executable plans, as they have been trained on a vast amount of code. One notable application is CLARify~\cite{yoshikawa2023large}, which utilizes GPT-3~\cite{brown2020language} to generate task plans in a specific Chemistry Description Language (XDL) based on descriptive user instructions in natural languages. Constrained task and motion planning problems are then solved using PDDLStream solvers. This approach aims to facilitate the autonomous and safe execution of chemistry experiments using general-purpose robot manipulators. Notably, these plans have shown much higher accuracy than baseline systems like SynthReader~\cite{mehr2020universal}.

Furthermore, there are preliminary attempts using GPT-4 to generate compatible scripts in Python to control OT-2~\cite{inagaki2023llms}, a computer-controlled liquid handling robot, achieving 95\% success within five iterations.

An emerging branch of AI research involves utilizing large language models as agents that will autonomously create, perform, and program scientific experiments. Boiko et al.~\cite{boiko2023emergent} presented one such method, showing how these models could use web search engines for information about molecule synthesis or employ vector search to find relevant documentation on chemical reactions. They have also developed multi-instrument systems code generation agents that can successfully implement complex experiments like Suzuki and Sonogashira cross-coupling reactions.

\textbf{Retrosynthetic Planning \& Reaction Prediction.}
Retrosynthetic planning involves breaking down complex compounds into simpler precursor compounds, while reaction prediction entails forecasting the outcome of chemical reactions. These two tasks are pivotal for understanding how to synthesize complex molecules from more basic starting materials, which is an essential step for preparing experiments like high-throughput screening. 

An early attempt at LLMs for retrosynthetic planning is the Molecular Transformer~\cite{schwaller2019molecular}, which utilizes a simple encoder-decoder transformer framework. The model is trained to take reactants and reagents as input and predict the chemical product that can be synthesized from a reaction. It has demonstrated higher accuracy in reaction prediction than human chemists. Subsequently, Schwaller et al.~\cite{schwaller2020predicting} combined the Molecular Transformer with a hyper-graph exploration strategy to develop an automated retrosynthetic route planning system. The dynamically constructed hypergraph represents a generic reaction where each molecule is a node, and the hyper-arc symbolizes the reaction arrow. 
This optimal synthetic route is identified using beam search over a beam search across the hyper-graph of possible disconnection strategies.

The capabilities of LLMs were further extended by Chemformer~\cite{irwin2022chemformer}, which is based on the BART~\cite{lewis-etal-2020-bart}  architecture. It includes an additional pretraining process that involves reconstructing masked SMILES strings and using an autoencoder to convert original SMILES to embeddings and back. The model is then fine-tuned for various downstream tasks, including reaction and retrosynthesis predictions. 

Recently, general-purpose LLMs have emerged in this field, such as Chemcrow~\cite{bran2023chemcrow} and Boiko et al~\cite{boiko2023emergent}. Boiko's system uses web search and simple calculations, while Chemcrow adopted a more sophisticated approach.
Chemcorw has developed and utilized a more comprehensive range of customized molecule and reaction tools. These include functionalities like converting queries to SMILES, obtaining molecule prices, patent checking, and reaction classification. Additionally, Chemcrow adopts a four-step framework to improve LLMs' ability: think about the necessary steps, take action using tools, provide inputs to these tools, analyze observations, and then deliver the final answer. This approach has been shown to perform better than GPT-4 in most tasks evaluated by humans in synthesis planning. Similarly, a recent study~\cite{jablonka2024leveraging} demonstrated that a fine-tuned GPT-3 model was shown to outperform traditional machine learning models on several chemistry tasks, especially in low-data scenarios. The findings highlighted that LLMs, even those not initially trained on chemical data, could adapt to various predictive chemistry tasks with minimal fine-tuning, showcasing the potential of LLMs in advancing chemical research.

\subsubsection{In-Silico Simulation}
In-silico simulation leverages computer models to simulate complex biological processes. These simulations are pivotal in understanding and predicting how drugs interact at the molecular level, leading to more efficient and targeted drug development. The three main tasks involved in in-silico simulations are de novo molecule generation, de novo protein generation, and protein-ligand interaction prediction.

\textbf{De novo Molecule Generation.}
De novo molecule generation is a complex task in in-silico simulations that involves creating new molecular structures with the potential to be effective drugs. This process is categorized into two types: unconstrained molecule generation, which seeks to populate the chemical space of the training set, and constrained molecule generation, where molecules are synthesized to meet specific desired properties~\cite{brown2019guacamol}. Constrained generation requires a model to consider various constraints such as affinity to targets, selectivity against off-targets, appropriate physicochemical properties, ADME characteristics, pharmacokinetics/pharmacodynamics, toxicology, and synthesizability~\cite{loeffler2023reinvent4}. 

To benchmark the performance of de novo molecule generation methods, GuacaMol~\cite{brown2019guacamol} was introduced to provide a benchmarking framework considering aspects such as validity, uniqueness, and novelty. Specialized language models have demonstrated remarkable proficiency. For instance, it has been have shown that even simple RNN-based models can perform exceedingly well on challenging generative modeling tasks~\cite{flam2022language}. These models effectively learn the complex distribution of molecules, such as the highest-scoring penalized logP molecules in ZINC15 or the most significant molecules in PubChem. To explore wide and novel chemical spaces, LLMs such as SMILES-LSTM and ORGAN~\cite{guimaraes2017objective} have been assessed with the unconstrained generation ability to directly generates~\cite{brown2019guacamol}, and ORGAN. When it comes to constrained molecule generation, several notable approaches have been developed. Previous studies utilized reinforcement learning~\cite{olivecrona2017molecular} and pharmacophoric features~\cite{skalic2019target} to improve RNN-based models toward generating molecules with desired properties and binding ligands to protein pockets.

Moreover, MolGPT~\cite{bagal2021molgpt}, with its GPT architecture, can handle multiple constraints. It is trained by recovering a molecule with its scaffold and properties. The REINVENT series~\cite{blaschke2020reinvent,loeffler2023reinvent4} represents a more advanced approach in this category. It is sweeping and capable of meeting up to 10 different objectives, including synthesizability, selectivity, etc. This method narrows the chemical search space through a three-staged training process: the prior model, a transfer learning agent, and staged learning towards generating high-scoring sequences. This sophisticated approach involves pretraining, transfer learning, reinforcement learning, and curriculum learning.

On the other hand, general LLMs usually focus on the constrained molecule generation task. In this context, MolT5~\cite{edwards2022translation} uses a self-supervised learning framework to pretrain T5~\cite{raffel2020exploring}, a general-purpose LLM that is trained on a large corpus of text coupled with molecular data pairs. The pretraining methodology comprises unsupervised SMILES recovery and associated chemical texts. More recently, GPT-4~\cite{gpt4-technical-report} has shown its ability to produce a novel molecule guided by textual instructions. However, the effectiveness generated by these two models is inferior to specialized language models. In addition, multimodal methods such as Momu~\cite{su2022molecular} and GIT-Mol~\cite{liu2023git} enhance general LLMs' capabilities in molecule generation. Momu~\cite{su2022molecular} improves upon MolT5~\cite{edwards2022translation} by adopting CLIP~\cite{radford2021learning} to align molecule graphs with related text. At the same time, GIT-Mol~\cite{liu2023git}, inspired by BLIP2's Q-FORMER strategy~\cite{li2023blip}, integrates graph, image and text information, using cross-attention and a variety of pretraining tasks. This multimodal approach significantly improves the effectiveness of MolT5~\cite{edwards2022translation} in constrained molecule generation tasks.

\textbf{De novo Protein Generation.}
Similar to molecule generation, this task focuses on designing new proteins in an unconditional manner~\cite{hesslow2022rita,ferruz2022protgpt2,nijkamp2023progen2} or conditional manner, where proteins generated should fit user constraints~\cite{wang2022scaffolding,ram2022few,watson2023novo}. Using these LLMs, scientists can design new artificial proteins that could serve specific functions, such as binding to a particular receptor or acting as enzymes.

Unconstrained generation aims to delve into and map the extensive protein space with specialized protein-based LLMs, such as those using autoregressive models for amino acid sequence generation, demonstrating remarkable effectiveness in this realm~\cite{madani2020progen,hesslow2022rita,nijkamp2023progen2}. A prime example of these LLMs is ProtGPT2, detailed in \citet{ferruz2022protgpt2}. Trained on a wide range of protein sequences, ProtGPT2 excels in creating de novo protein sequences that mirror natural patterns. Its outputs, marked by ordinary amino acid propensities, are predominantly globular, resembling natural proteins. Intriguingly, comparative analysis with protein databases indicates that ProtGPT2's generated sequences distantly related to existing proteins can form valid structures based on AlphaFold2. This indicates ProtGPT2's ability to explore novel and valid protein space areas.

Constrained protein generation seeks to achieve the controllable design of novel proteins with specified cellular compartments or functions~\cite{ferruz2022controllable}. A typical practical constraint is generating protein sequences within the same family. This method ensures that new proteins retain the critical characteristics of a given group. Specialized LLMs like ProGen~\cite{madani2023large} and PoET~\cite{watson2023novo} are utilized in this context. ProGen specifically generates sequences for a particular family, using a prefix like ``Protein Family: Pfam ID'', and has produced proteins with efficiencies close to natural counterparts, even with low sequence identity. PoET, on the other hand, compiles multiple sequences from the same family to create new sequences, akin to forming a paragraph from multiple sentences, thereby preserving the family's structural integrity. 

Building on this, a more challenging yet direct path to drug development in constrained protein generation is the design of protein binders~\cite{mcpartlon2023deep}. This task requires intricately designing proteins for specific binding functions, necessitating a deep understanding of how the entire protein folds into a desired structure with a few functional residues underpinned by the protein's overall structure. A more streamlined approach involves starting with a desired structure and, conditional on this structure, using a specialized LLM for inverse folding~\cite{hsu2022learning}. This inverse folding method starts with a complex protein structure and only seeks to leverage LLMs to convert the desired structure into protein sequences. 

\citet{watson2023novo} modified RoseTTAFold, a specialized LLM already equipped with profound structural knowledge and an understanding of protein folding, with diffusion modeling to create RFDiffusion. This innovative approach allows RFDiffusion to begin with randomly distributed residue frames and systematically denoise them toward a valid protein structure. More recently, the development of RFdiffusion All-Atom (RFdiffusionAA)~\cite{krishna2024generalized} extends this capability by integrating atomic-level detail to generate folded protein structures that specifically bind to small molecules, metals, and nucleic acids. By initializing the model with random distributions of residues around these small molecules, RFdiffusionAA can design highly specific binding pockets that are validated both computationally and experimentally. Subsequently, RFDiffusion can restructure the sequence in an inverse folding manner for greater accuracy, facilitating the design of diverse functional proteins based on simple molecular specifications. This method complements and enhances the potential of RoseTTAFold by enabling not only constrained protein design but also the creation of entirely new protein structures around various molecular targets. Notably, most of these constrained-based LLMs can also generate unconstrained protein structures, either by iteratively exploring the constrained space or by initializing the model via a random process.

Specialized protein-based LLMs above have shown considerable promise; however, despite evidence that they can design protein sequences that align with user requirements, it remains to be seen whether these LLMs, trained on natural protein sequences, can generalize beyond these to unnatural sequences. To address this question, \citet{verkuil2022language} leveraged the ESM protein language model~\cite{rives2021biological} to design novel protein structures in unconstrained and constrained scenarios. By experimentally validating the generated proteins, it achieved a 67\% success rate in creating functional proteins, some of which bear minimal similarity to known proteins. The impact of this validation on drug discovery and development is significant: it substantiates innovative LLM approaches for creating biologics and therapeutic proteins, potentially speeding up the creation of new protein-based treatments. Notably, with the advent of ESM3~\cite{hayes2024simulating}, the newer version of ESM, improvements in the ability to predict and design complex protein structures have been demonstrated. 

More recently, general-purpose LLMs represent a newly emerging de novo protein generation technology. ProteinDT~\cite{liu2023text}, a multi-modal framework, incorporates textual descriptions for protein design. To train ProteinDT, a dataset of 441K text and protein pairs was constructed. By training on these pairs, ProteinDT was able to achieve over 90\% accuracy for text-guided de novo protein generation. Furthermore, ProteinDT can be used to perform text-guided
protein optimization tasks, as discussed in section 3.2.4. 

\textbf{Protein-ligand Interaction Prediction.}
Understanding the mechanisms underlying the interaction between a drug (ligand) and its protein target is fundamental to drug discovery and development. In recent years, virtual screening via in silico methods such as molecular docking or classic predictive machine learning models has been at the forefront of streamlining drug development processes. These tools form the cornerstone for accelerating the early-stage identification of potential therapeutic agents. In the early stages of drug discovery, specialized large language models (LLMs) are utilized in two primary directions: (i) to use LLM directly as a backbone and (ii) as a standalone yet essential part of a more comprehensive predictive system. Both LLM use cases are highly instrumental in improving the efficiency and effectiveness of the drug development process. 

Tools such as AlphaFold-multimer~\cite{evans2021protein}, used explicitly for protein-protein docking (as detailed in section 3.1.3 on protein-protein interaction), highlight the direct application of specialized LLMs as a backbone. Likewise, \citet{singh2023contrastive} have employed protein LLM alongside molecular fingerprints for virtual screening. This technique has successfully identified binders with sub-nanomolar affinity, underscoring LLMs' increasing impact and potential in refining the drug discovery and development process. 

More commonly, incorporating protein or ligand embeddings and structural information from specialized LLMs as part of complex system architectures has become a standard in assisting accurate prediction and docking~\cite{wang2022structure,lu2022tankbind, Jiang2022SequencebasedDA,corso2023diffdock,ketata2023diffdockpp,koh2024psichic}. PSICHIC~\cite{koh2024psichic} demonstrated that learning from sequence data alone (protein sequence and ligand SMILES string) can surpass methods that rely on experimental 3D structures or protein-ligand complexes. More significantly,  when learning from sequence data alone, PSICHIC demonstrated emergent capabilities in deciphering the mechanisms underlying protein-ligand interactions. It successfully identifies protein residues in the binding site and ligand atoms involved in these interactions, a capability achieved through training PSICHIC exclusively on sequence data without any information on specific binding sites or residue-atom interactions. Such achievements suggest a highly promising avenue for LLMs in processing extensive biochemical data. If successful, this approach can enable prediction and foster an understanding of how various chemical structures interact with different proteins, potentially revealing many hitherto unknown aspects of the protein-ligand interactions without costly experimental endeavors.

In general-purpose Large LLMs, Galactica is equipped with in-depth general scientific knowledge and has also been jointly trained to predict the docking scores of protein-ligand sequences~\cite{taylor2022galactica}. It has demonstrated reasonable correlation with experimental results in certain instances. It is intriguing to consider the potential of developing an LLM that can directly understand, interpret, and predict protein-ligand interactions at a molecular level while also encompassing general knowledge.  

\subsubsection{ADMET Prediction}
The prediction of absorption, distribution, metabolism, excretion, and toxicity (ADMET) attributes of compounds is a critical phase during the "Hit to Lead" and "Lead Optimization" stages of drug development. This task is essential to distinguish compounds with favorable pharmacokinetic profiles from those with negative characteristics, ensuring the progression of only the most promising drug candidates. Predicting molecular properties for multiple scientific areas, including physiology, physical chemistry, biophysics, and quantum mechanics, is the core of molecular property prediction in drug discovery. In recent years, both specialized and general-purpose large language models (LLMs) have shown remarkable predictive capabilities in the field of ADMET prediction, leveraging their ability to learn large amounts of data.

Specialized LLMs are usually trained on large datasets of SMILES strings and then fine-tuned for specific downstream property prediction tasks. For instance, ChemBERTa, based on the extensive PubChem 77M dataset, has shown comparable performance with traditional machine learning approaches like random forests and support vector machines. Similarly, SMILES Transformer, which employs an encoder-decoder architecture trained on over 861,000 unlabeled SMILES strings, has reached state-of-the-art performance. Recently, large-scale transformer-based models such as Molformer and BARTSMILES have demonstrated new state-of-the-art performance, though they require vast training time and resources.

General LLMs, on the other hand, either find knowledge to augment traditional machine learning models or are fine-tuned for specific tasks. LLM4SD can synthesize rules from literature and data sources that allow even random forest models to outperform all state-of-the-art methods for most tasks. Galactica and other models pre-trained on a significant amount of scientific literature have demonstrated capabilities in molecular property prediction with simple text instructions. Additionally, previous studies have highlighted the potential of other models such as GIT-Mol and MolT5, which could also yield satisfactory results after being fine-tuned on downstream tasks. GPT-4's extensive training on diverse text data enables it to provide valuable insights in the field, although adapting it for specific molecular property predictions might be challenging due to its proprietary nature.

\subsubsection{Lead Optimization}
Lead optimization is a significant step that aims to modify the drug candidate molecular structure or protein sequence to enhance its potency, safety, and stability. This process is usually carried out by chemists or biologists who modify according to their knowledge and experience. However, this process is time-consuming and requires much effort because it may take several attempts before arriving at the desired outcome. LLMs can assist with this task by utilizing statistical analysis on large data sets to predict how altering the structure of a compound would affect its properties. This feature supports more effective decision-making for chemists, minimizing the number of trials required for optimizing compounds.

\textbf{Molecular Optimization.}
Molecular optimization is a complex task that involves modifying a molecular compound's structure to enhance its efficacy, stability, and safety. This process can be divided into two major categories: uncontrolled and controlled. In uncontrolled optimization, the core scaffold will be perserved but the LLMs will randomly modify the surrounding functional groups with external guidance to improve property values. In contrast, controlled optimization means users can specify parts of a molecule to be optimized for a given property. Through editing, the final compounds are expected to have enhanced properties. Compared with uncontrolled approaches, these methods allow chemists to have a greater degree of specification.

Specialized LLMs can aid in molecular optimization through both uncontrollable and controllable strategies. For uncontrollable optimization, models like the Reinvent series \cite{blaschke2020reinvent, loeffler2023reinvent4} and MERMAID \cite{erikawa2021mermaid} use reinforcement learning to ensure that the synthesized molecules retain the desired structural scaffolds while enhancing properties such as potency, stability, or drug-likeness, guided by external models like drug-likeness filters or predictive algorithms. Specifically, MERMAID~\cite{erikawa2021mermaid} incorporates a Monte Carlo Tree Search (MCTS) strategy, adeptly navigating potential molecular modifications to find the best ones. Apart from reinforcement learning, alternative methods such as fine-tuning and pretraining are also employed to enable specialized LMs with molecular optimization capabilities. An example is LigGPT~\cite{bagal2021liggpt}, which focuses on generating molecules with specific scaffolds and desired properties. This model uses a trained transformer decoder architecture to reconstruct the original SMILES strings with the given scaffold and property information. For controllable optimization, Transformer-R~\cite{he2021transformer} and He et al.~\cite{he2021molecular} demonstrate its molecular optimization capability based on matched molecular pair (MMP) analysis. This model successfully considers property constraints and crucial components of SMILES strings while producing the required source R-GROUP for targeted molecular optimization. Improving these two approaches, C5T5~\cite{rothchild2021c5t5} does not rely on molecular pair data for training. This method is trained to recover masked IUPAC names by using specific tokens denoting the property range of the molecule. It shows successful applications for logP, logD, PSA, and Refractivity among other properties.

Recently, general LLMs have emerged, which present a novel avenue for amalgamating human expertise with LLM capability. The first one to consider is MoleculeSTM~\cite{liu2022multi}, an application of multimodal learning for learning molecular structures from textual descriptions through contrastive learning. A dual-phase strategy is employed to enable a generative model to perform molecule editing. The first step involves training an adapter that aligns a molecule generative model with a joint representation space. In contrast, the second step focuses on fine-tuning the latent space to minimize differences between the generated molecule and given molecule-text instructions. An approach called ChatDrug~\cite{liu2023chatgpt} significantly outperforms MoleculeSTM~\cite{liu2022multi} attributed to its integration with ChatGPT, which provides conversational capabilities. This agent-driven technique applies an iterative refinement procedure combining information retrieval and domain-specific feedback, improving drug optimization workflow. ChatDrug incorporates modules like the Prompt Design for Domain-Specific (PDDS) module, which uses extensive prompt engineering from language models like LLMs. Later on, the Retrieval and Domain Feedback (ReDF) module will help find molecules based on specific requirements.

Additionally, GPT-4 has demonstrated basic abilities to optimize molecular structures, even though it was not explicitly developed for this purpose~\cite{gpt4-technical-report}. However, there are certain limitations; GPT-4 can innovate upon existing compounds without incremental feedback and continuous checks from humans. It can easily lead to inaccurate responses.

\textbf{Protein Optimization.}
Like molecular optimization, protein optimization involves modifying the structure to enhance protein functionality and safety. In this field, biochemists and molecular biologists meticulously adjust proteins, a process that can be laborious and iterative. LLMs can contribute by offering predictions on how structural changes impact protein properties. Specifically, in the development of antibody drugs, language models can be used to consider multiple essential factors, including improving antigen binding, reducing immunogenicity, enhancing stability, and preventing high viscosity or polyspecificity~\cite{beck2017strategies,nichols2015rational,raybould2019five}.

In uncontrolled optimization, ESM~\cite{rives2021biological}, which is a protein-based LLM trained on diverse protein sequences with general evolutionary information, has been used to suggest evolutionarily viable mutations that could help enhance fitness across protein families~\cite{hie2023efficient}. This work relies on evolutionary plausible mutation that generally improve fitness across proteins rather than specific properties. As a proof-of-concept, this work found notable improvements in matured IgG antibodies' affinity against different viral antigens achieved with minimal testing of the variants through only two rounds of evolution.

In controlled optimization, protein hallucination and inpainting have been proposed~\cite{wang2022scaffolding} that resembles constrained optimization have been proposed to optimize proteins using a section of the protein masked off, and a protein LLM~\cite{dauparas2022robust} is used to sample and refine the protein sequence while maintaining the original backbone structure. 

Among the general-purpose LLMs, ProteinDT~\cite{liu2023text} is a novel method that optimize protein sequences using prompts encapsulating specific properties information. ProteinDT can understand both natural texts and protein sequences.
Leveraging this ability, the method employs two techniques: latent interpolation and latent optimization. Latent interpolation merges text prompt and protein sequence representations, while latent optimization aligns which aligns a token-level latent code with both text and protein representations. Experiments showcase that LLMs can conduct protein optimization, including structure, stability, and peptide binding optimization. Adopting such a strategy indicates a novel way in protein engineering for achieving text-guided, exact transformations on protein structures aiming at their required features.

\subsection{Assistance}
In the same way that researchers in the "Understanding Disease Mechanisms" phase require a diverse array of information sources, those involved in drug discovery also need access to various relevant resources. These resources can include, but are not limited to, comprehensive compound libraries, up-to-date research publications, and extensive patent landscapes.

To gather this information, a procedure known as information retrieval is used, which involves using General LLMs with web searching and knowledge retrieval abilities to source information from various sources, such as research articles, compound databases, and patent documents. Additionally, tools like Galactica~\cite{taylor2022galactica} and GPT4~\cite{ai4science2023impact} can assist in clarifying scientific concepts and data, helping researchers achieve a deeper understanding.

\subsection{Clinical Trials}
Clinical trials represent the last stage of the drug development pipeline and are essential to evaluate a drug candidate's safety and efficiency. These trials take place in four phases, each serving a specific purpose. Phase 1 involves a small number of healthy volunteers who are administered the compound to evaluate its safety and tolerability. Phase 2 involves a larger group of patients to evaluate efficacy and side effects. Phase 3 is the last testing stage performed on many patients after an optimal treatment selection. In this phase, the new treatment is compared to existing treatments to understand its differences. The final phase is postmarketing surveillance, during which the drug is monitored for adverse effects. 

Using LLMs, one of the most significant areas where they can be used is clinical practice, result analysis, and clinical assistance. The subsequent paragraphs will delineate the specific downstream tasks for each category of these tasks and how LLMs can be incorporated into them to enhance drug discovery further.

\subsubsection{Clinical Practice}
In the realm of clinical trials, practitioners are typically tasked with four core responsibilities: coding ICD, matching patients with trials, predicting outcomes, and planning the trials themselves. These responsibilities span various areas and have traditionally been fulfilled by experienced practitioners who rely on their knowledge and expertise. Hence, clinical trial practitioners face the challenge of reading and understanding large amounts of information such as electronic health records (EHRs), trial eligibility criteria (ECs), trial protocols, and outcome reports. Fortunately, general LLMs have emerged as a promising solution for accelerating these processes, as they excel at extracting and handling information from large volumes of text data.

\textbf{ICD Coding.}
ICD coding is an essential practice in clinical practice, which deals with assigning ICD codes to patient records. This is usually a time-consuming and laborious process. By analyzing vast amounts of EHR data, LLMs can predict the most suitable codes for Electronic Health Records (EHRs), thus enabling clinical practitioners to make more enlightened decisions and streamline the process.

Shi et al. created a groundbreaking system~\cite{shi2017towards} that utilizes a character-aware LSTM-based network to process diagnosis descriptions from hospital admission records more efficiently. Similarly, Xie and Xin~\cite{xie2018neural} developed a tree-of-sequences LSTM network to encode diagnosis descriptions and ICD codes. Inspired by Shi et al.'s work~\cite{shi2017towards}, they added adversarial learning to align various writing styles using an attentional matching module with isotonic constraints. These enhancements improve code assignments and prioritize more significant codes.

Some recent studies have used more contemporary and up-to-date frameworks for their language models. For example, BERT-XML~\cite{zhang-etal-2020-bert} incorporates BERT pretraining with multi-label attention to encode EHRs and later uses a multi-label classification model to predict ICD codes from EHRs. This approach takes advantage of the capabilities of the BERT framework to increase the accuracy of code assignment. Another example, PLM-ICD~\cite{huang-etal-2022-plm} adapts domain-specific pretrained language models such as BioBERT~\cite{lee2020biobert}, PubMedBERT~\cite{gu2021domain}, and RobBERTa-PM~\cite{liu2019roberta}, for ICD coding by fine-tuning them. It also uses segment pooling and label attention to increase efficiency and accuracy in clinical coding contexts.

\textbf{Patient-Trial Matching.}
When it comes to matching patients with clinical trials, the process relies on the use of electronic health records (EHRs) to identify viable options based on the patient's medical history. Historically, this task was performed manually by physicians and data analysts who would sift through patient demographics and pre-screening eligibility factors to pinpoint the most suitable trial. However, this approach can be time-consuming and fraught with errors due to the complexity and diversity of trial criteria.

To overcome these challenges, preliminary works typically encode EHRs and eligibility criteria into an embedding pair and then calculate the match score through similarity. An example of this is the cross-modal framework called DeepEnroll~\cite{zhang2020deepenroll}, which utilizes BERT to capture eligibility criteria from the text-based patient records, using a hierarchical structure of latent representation. This framework finally observed a notable 12.4\% 

COMPOSE~\cite{gao2020compose} significantly improves DeepEnroll by utilizing a dual pathway encoding framework and a composite loss function. This approach effectively separates inclusion and exclusion criteria. The ECs pathway employs BERT and convolutional neural networks (CNNs) for robust encoding, while in the EHR pathway, hierarchical memory networks are deployed to organize medical concept hierarchies systematically. COMPOSE~\cite{gao2020compose} then dynamically interacts with these memories using EC embeddings, which help it to choose the most precise matches. 

Recently, there have been several methods harnessing general-purpose LLMs to facilitate patient-trial matching based on LLMs reasoning ability. Med-monoT5~\cite{pradeep2022neural} is a T5-based system fine-tuned on medical passage ranking tasks that follows a zero-shot approach. It evaluates clinical trial documents' relevance to patient descriptions utilizing specifically designed templates. It employs a two-stage fine-tuning process on general and medical datasets, leveraging a sliding-window approach to handle lengthy text fields for matching patients with appropriate clinical trials. Hamer et al. ~\cite{hamer2023improving} use InstructGPT~\cite{ouyang2022training}  to assist physicians in determining patient eligibility for clinical trials. Employ prompting strategies such as one-shot, selection-inference, and chain-of-thought—to parse and analyze the criteria. While this automation has been shown to potentially reduce up to 90\% of the workload, achieving about 72\% accuracy in screenability, it is not without issues. Overconfidence in interpreting ambiguous criteria and the risk of generating inaccurate content necessitate continued supervision by medical professionals to ensure reliability. Another pioneering work, TrialGPT~\cite{jin2023matching}, uses an architecture that predicts criterion-level eligibility and provides detailed explanations. These explanations are aggregated to rank and exclude candidate clinical trials based on free-text patient notes. Although TrialGPT~\cite{jin2023matching} correlates well with expert annotations, its occasional errors highlight the limited medical knowledge of GPT 3.5 and the need for their careful integration into clinical trial matching processes.

\textbf{Clinical Trial Planning and Prediction.}
Planning a clinical trial involves several labor-intensive activities, including searching for historical trials, designing trial criteria, and selecting suitable sites. To accelerate this process, general LLMs have been introduced in the domain. These models utilize historical data on trials, patient outcomes, and demographic trends to provide informed suggestions.

The first step in preparing a clinical trial is the search for historical trials, simplified by embedding trials into documents using document embedding methods. Although some traditional statistical models like BM25~\cite{trotman2014improvements} and TF-IDF~\cite{ramos2003using} can produce such embeddings, their performance often needs to catch up to that of deep learning. Deep learning models such as Doc2Vec~\cite{le2014distributed} and BERT~\cite{devlin-etal-2019-bert} provide more sophisticated embeddings but still face challenges in medical-specific retrieval tasks. These gaps are narrowed by medically tailored adaptations of BERT~\cite{devlin-etal-2019-bert}, including clinical BERT~\cite{alsentzer2019publicly}, clinical bioBERT~\cite{alsentzer2019publicly}, TrialBERT~\cite{wang2022trial2vec}, and Med-monoT5~\cite{pradeep2022neural}, which have been fine-tuned on substantial medical literature to provide highly accurate retrieval results. Going beyond these developments, the Trial2Vec ~\cite{wang2022trial2vec} framework extends from TrialBERT ~\cite{wang2022trial2vec} and utilizes hierarchical contrastive learning to generate global and local embeddings that incorporate semantic meaning based on document meta-structure. Additionally, cliniDigest~\cite{white2023clinidigest} employs the summarization capabilities of GPT-3.5 to provide up-to-date, concise summaries of clinical trials, thereby facilitating quick decision-making for upcoming trials.

AutoTrial utilizes a two-stage training method with GPT-2 as its backbone to automate trial criteria design and clinical trial planning. This model employs the decoder architecture, learns from vast documents regarding previous trials during pretraining, and then gets its task-specific fine-tuning to generate exact trial criteria from given specifications. Its combination of a hybrid prompting strategy and multi-stage training makes it easy to adapt without retraining or performance loss.

Trial site matching involves finding a suitable trial site for a clinical trial. Modern algorithms integrate multimodal data containing unstructured and structured information about a trial to rank sites. One such algorithm is PG-Entropy~\cite{srinivasa2022clinical}, which exploits ClinicalBERT~\cite{huang2019clinicalbert}-based encoding of trial criteria text in combination with structured clinical trial features via a list-wise policy learning approach. It also gives equal importance to ensuring sensitive attributes do not adversely impact any bias-related outcome. In contrast, FRAMM~\cite{theodorou2023framm} uses a deep reinforcement learning mechanism.

In this way, it is an effective remedy for the absence of data and for creating site representations using the masked cross-attention mechanism. In addition, it balances diversity and enrollment when choosing sites via a Deep Q-Value Network that helps in decision-making by defining the reward function that focuses on both these aspects.

To predict the success of a clinical trial, trial outcome prediction analyzes various trial-related variables, such as information about the disease, the drug, trial criteria, and trial protocol. LLMs can provide insightful predictions to help trial designers enhance the trial plan. This can mitigate the risk of adverse outcomes from suboptimal trial configurations.

Existing methods mainly use LLMs to encode clinical trial information, which is then used to predict clinical trial outcomes. These LLM-based methods performed better than traditional machine learning methods such as logistic regression, MLP, and XGBoost~\cite{chen2016xgboost}. For instance, given drug, disease, and clinical protocol information, HINT~\cite{fu2022hint} amalgamates comprehensive web data such as drug properties, disease characteristics, and clinical trial information to assist the model in predicting both phase-level and trial-level clinical trial outcomes. Specifically, it separately encodes the drug molecules, disease characteristics, and clinical protocols using pre-trained models with web knowledge. Then, it builds an interaction graph using these components to generate the final prediction. 

SPOT~\cite{wang2023spot} and HINT~\cite{fu2022hint} are both methods that aggregate clinical trials of the same topic into a sequence based on timestamps instead of making predictions for each trial individually. SPOT uses a sequenced meta-learning approach that begins with topic discovery and clustering, followed by an RNN structure for sequential predictive modeling. It then uses a meta-learning approach to optimize models for different topics. On the other hand, MediTab~\cite{wang2023meditab} can be adapted to perform clinical trial prediction and achieves better performance than both HINT and SPOT, with the added advantage of using larger-scale tabular data from different sources. While DeepEnroll~\cite{zhang2020deepenroll} and COMPOSE~\cite{gao2020compose} were initially designed for patient-trial matching, they can also be adapted for clinical trial prediction. However, they perform less effectively than previous LLM methods.

\textbf{Documents Writing.}
The generation of various clinical documents, such as discharge summaries, clinical notes, and radiology reports, has traditionally been a time-consuming and laborious task performed by clinical practitioners. However, language models (LLMs) are increasingly being employed to automate this process with their powerful text-generation ability. 

For instance, in discharge summaries generation, Patel et al.~\cite{patel2023chatgpt} present a prompt to generate discharge using chatGPT automatically, while Shing et al.~\cite{shing2021towards} have implemented an extractive-abstractive summarization pipeline. This method consists of a two-stage process that begins by extracting relevant sentences from clinical notes, followed by an abstractive summarization technique to transform these extracts into coherent summaries.

LLMs have also been shown to be highly effective in maintaining regular check-ins with patients, as demonstrated by ~\cite{webster2023six}. These models can assist in writing notes in patients' records and summarizing their issues, bridging the gap between conventional encounters with caregivers. In addition, Seppo et al.~\cite{enarvi-etal-2020-generating} generated medical reports from patient-doctor conversation transcripts using RNN and transformer models. To achieve this, they trained train RNN and transformer models using a dataset of around 800,000 orthopedic encounters. When both RNN and Transformer models were compared, the latter showed superior accuracy and training efficiency performance. 

For generating reports from randomized controlled trials (RCTs), the RobotReviewer~\cite{marshall2017automating} system has shown its capability to automatically generate reports summarizing critical information from RCTs. 

Finally, multi-modal LLMs have been widely used for the task of radiology report generation. In the early days, specific language models were employed as encoders. An example is TieNet~\cite{wang2018tienet}, an end-to-end CNN-RNN architecture with multi-level attention models. It involves merging image features and text embeddings from associated reports to improve disease classification accuracy and report quality~\cite{wang2018tienet}. Taking a step further, Liu et al.~\cite{liu2019clinically} presented a hierarchical generation strategy via CNN-RNN-RNN architecture with reinforcement learning. The approach focuses on balancing readability and clinical accuracy by considering Clinically Coherent Reward (CCR) to maintain the clinical relevance of the reports. For methods using more modern LLMs, MedViLL~\cite{moon2022multi}, which also uses the BERT~\cite{devlin-etal-2019-bert} architecture, concatenates visual and language feature embeddings and trains them on tasks like masked language modeling and image report matching to achieve an effective alignment of visual and language features. Another example, RadBERT~\cite{yan2022radbert}, is a BERT-like system pre-trained on millions of radiology reports that can generate concise reports highlighting essential observations and conclusions. More recently, during human evaluation, Med-PaLM M~\cite{tu2024towards}, a multimodal medical language model, generated chest X-ray reports that clinicians preferred over those by radiologists in up to 40.50\% of cases.

\subsubsection{Patient Results}
LLMs can help in predicting patient outcomes using patient visits data. There are two categories of downstream tasks in this domain: hospital-related and disease-related predictions. While the former type includes hospital readmission, length of stay, and mortality, the latter type focuses on disease onset, diagnosis, and morbidity.

\textbf{Patient Outcome Prediction.}
Patient outcome prediction aims to predict a patient's current or future health status. Language models help in this area by encoding vast amounts of electronic health records (EHRs) to predict future health outcomes, thereby enabling clinicians to make more informed decisions and potentially saving a substantial amount of time and resources. These LLMs-based methods generally achieved better performance than traditional machine learning methods such as logistic regression, MLP, and XGBoost~\cite{chen2016xgboost}.

There are LLMs focusing on hospital-related predictions. For example, ClinicalBERT~\cite{huang2019clinicalbert} leverages the BERT~\cite{devlin-etal-2019-bert} architecture to understand clinical records and make readmission predictions. Similarly, NYUTron~\cite{jiang2023health}, a BERT-like language model, is pre-trained on a comprehensive collection of clinical notes and fine-tuned for various tasks such as predicting mortality, comorbidity, hospital readmission, insurance denial, and length of stay. Another model, RAIM~\cite{xu2018raim}, processes multimodal EHR data—including clinical records, electrocardiograph waveforms, and vital signs—to predict outcomes like decompensation and length of stay. It considers both historical and current data points, focusing on the most relevant information for accurate predictions. StageNet~\cite{gao2020stagenet}, which can predict decompensation and mortality, uses a stage-aware LSTM module that captures changes in patients' health conditions over time. It integrates time information between visits and employs a stage-adaptive convolutional module to recalibrate understanding of disease progression, highlighting the most informative patterns for precise outcome prediction.

The second category of methods focuses on disease-related predictions, including disease onset, diagnosis, and morbidity. For, instance, RETAIN~\cite{choi2016retain} employs a two-level neural attention mechanism with recurrent neural networks (RNNs) to focus on crucial visits and variables for heart failure predictions. Building upon RETAIN, Dipole~\cite{ma2017dipole} uses a bidirectional RNN and three types of attention mechanisms to enhance diagnosis predictions, determining future medical codes for patient visits. With more recent LLMs, MediTab~\cite{wang2023meditab} consolidate and align different types of medical data, predicting patient outcomes such as morbidity. 

\subsubsection{Assistance}
General-purpose language models (LLMs) can play a significant role in clinical trial assistance by helping patients understand trial-related information, assisting clinicians in retrieving and understanding relevant literature and patient data, and supporting pharmacovigilance efforts by identifying and reporting adverse events.

General-purpose LLMs, such as GPT-4, and Med-Palm2 are capable of understanding medical knowledge and explaining it in simple, accessible language~\cite{kung2023performance, thirunavukarasu2023large}. This ability can help patients better comprehend and participate in clinical trial opportunities. Clinicians can also utilize LLMs to efficiently retrieve relevant clinical trial literature and assess patient eligibility using advanced information retrieval capabilities. In pharmacovigilance, LLMs contribute to understanding drug-drug interactions~\cite{luo2022biogpt,taylor2022galactica}, providing deeper insights into drug safety. Their code generation capabilities streamline data analysis~\cite{gpt4-technical-report}, enhancing the efficiency and speed of data interpretation.

\section{LLMs Maturity Assessment}
In this analysis, we evaluate the progress of two LLMs paradigms across 14 downstream task categories within three stages of the drug discovery and development pipeline: understanding disease mechanisms, drug discovery, and clinical trials. We categorize the maturity of these tasks using a four-tiered system ranging from ``Not Applicable'' to ``Matured''. We begin by outlining the specifics of the criteria before detailing the maturity levels of the various tasks within each phase.

\subsection{Maturity Criteria}
The maturity of LLMs in the drug discovery and development pipeline is categorized into four distinct levels (Figure~\ref{fig:maturity_assessment}):

\begin{enumerate}
    \item \textbf{\textit{Not Applicable:}} The LLM paradigm is not suitable or relevant for the given downstream task.
    \item \textbf{\textit{Nascent:}} The LLM paradigm has been applied to the task in a preliminary, in silico setting only and lacks validation through real-world experiments.
    \item \textbf{\textit{Advanced:}} The LLM paradigm has moved beyond theoretical application, with its effectiveness validated through real-world experiments in relevant scenarios.
    \item \textbf{\textit{Matured:}} The LLM paradigm application has been integrated and deployed in practical, real-world environments such as hospitals or pharmaceutical companies, with evidence demonstrating its effectiveness and utility.
\end{enumerate}

\subsection{Maturity Assessment of Downstream Tasks}
For the maturity assessment, we consider 14 different downstream tasks to be carried out in three significant drug discovery and development stages. In understanding disease mechanisms, we focus on genomics analysis, transcriptomic analysis, protein target analysis, disease pathway analysis, and assistance. At the stage of drug discovery, we analyze chemistry experiments, in silico simulations, ADMET prediction, lead optimization, and assistance. Finally, clinical trials deal with clinical applications in clinical practices, patient results, and assistance.

\subsubsection{Understanding Disease Mechanism}
\textbf{Genomics Analysis.} 
Specialized LLMs have been recently created to encode information in the nucleotide sequences~\cite{ji2021dnabert, dalla2023nucleotide, li2023applications} with any practical applications, such as genetic variant analysis~\cite{le2022bert, zhou2023dnabert}. However, they are still nascent, and further experiments are required to validate their effectiveness.

Recently, general LLMs have emerged and are also in the nascent stage. There is still room for improvement in explaining the evolutionary processes of genomic data or designing DNA sequences, as indicated in recent studies~\cite{ai4science2023impact}. This underscores the ever-developing field with its enormous potential for future breakthroughs.

\textbf{Transcriptomics Analysis.} There have been significant advancements in the real-world applications of specialized LLMs, such as the Geneformer LLM~\cite{theodoris2023transfer}. This particular LLM has played an essential role in gene network analysis. It was able to distinguish between normal and cardiomyopathic cardiomyocytes. This process identified the genes responsible for network perturbations associated with hypertrophic and dilated cardiomyopathy, providing potential therapeutic targets like ADCY5 and SRPK3. The real-world effectiveness of these targets was confirmed through experimental validation using iPSC-derived cardiac microtissues with Titin truncating mutations~\cite{theodoris2023transfer}. Hence, it is evident that specialized LLMs are in an advanced stage of development for analyzing transcriptomic data and deciphering disease mechanisms.

In contrast, general LLMs are still in the early nascent stage of development for transcriptomic analysis. Research is underway to explore auxiliary tasks in this field, such as automating cell type analysis~\cite{hou2023reference} and analyzing data through code generation~\cite{ai4science2023impact}.

\textbf{Protein Target Analysis.} Specialized LLMs for protein target analysis have significantly matured following the breakthrough of AlphaFold2~\cite{jumper2021highly}. AlphaFold2 is now a comprehensive and readily accessible database~\cite{varadi2022alphafold}, with various applications in structure-based drug discovery and vaccine development~\cite{varadi2023impact}. One noteworthy accomplishment is the rapid development of a first-in-class hit molecule for a novel target, CDK20, without an experimental structure, achieved within 30 days from target selection and requiring the synthesis of only seven compounds~\cite{ren2023alphafold}. Additionally, ESM~\cite{rives2021biological}, a prominent protein language model, has been developed into a web server focusing on GPCR proteins. This tool analyzes their signaling and functional repertoire~\cite{matic2022precogx} and identifies compounds with subnanomolar affinity~\cite{singhal2023large}.

While general LLMs like GPT-4~\cite{gpt4-technical-report} have demonstrated potential in analyzing complex scientific data, including protein analysis, they remain in a nascent stage of development within protein target analysis. ProteinChat~\cite{guo2023proteinchat} is another example demonstrating how these models can label protein structures based on user prompts. However, these developments primarily focus on generating informative answers based on embeddings without extensive real-world validation. Thus, this indicates that the field is still evolving.

\textbf{Disease-pathway analysis.} We are witnessing specialized LLMs reach an advanced stage in disease pathway analysis. These specialized LLMs have made significant contributions, particularly in genomics, transcriptomics, and protein target analysis. Notably, these fields play a critical role in the comprehensive analysis of disease pathways. A notable recent breakthrough is the use of transcriptomic LLM, Geneformer~\cite{theodoris2023transfer}, for gene network analysis, which has undergone laboratory validations and illustrates the capabilities of these models in dissecting disease pathways~\cite{theodoris2023transfer}.

General LLMs have also reached an advanced stage in disease-pathway analysis. For instance, Insilico Medicine, a biotechnology company, already offers ChatGPT integration with its PandaOmics target discovery platform to analyze disease pathways~\cite{savage2023drug}. While the PandaOmics target discovery already incorporated general LLMs for this purpose, these tools' widespread adoption and broad usage still need to be fully realized.

\textbf{Assistance.} General LLMs have attained a mature development stage, greatly aiding researchers in information retrieval and knowledge discovery for the understanding of disease mechanisms~\cite{ai4science2023impact,savage2023drug,toufiq2023harnessing}. These models possess exceptional capabilities in mining and synthesizing extensive scientific and medical literature, allowing us to understand these mechanisms better. Additionally, their skill in creating and interpreting knowledge graphs~\cite{savage2023drug} is crucial in mapping gene networks and elucidating gene-disease relationships~\cite{luo2022biogpt}. Additionally, general LLMs have proven effective in simplifying complex medical and genetic concepts~\cite{jeblick2023chatgpt}, thus making technical knowledge more accessible and enhancing education and communication in the medical field.

\subsubsection{Drug Discovery}
\textbf{Chemistry Experiments.}
For chemistry experiments, while the utilization of specialized LLMs is still in its nascent stages, general LLMs have advanced considerably. They are now used in sophisticated chemistry experiments (Figure ~\ref{fig:maturity_assessment}). Specialized LLMs, typically conducted in silico, are considered inferior to their general counterparts based on their performance in retrosynthetic planning and reaction prediction~\cite{boiko2023emergent, bran2023chemcrow}. This is mainly due to the tool use capabilities of general LLMs, which include using tools, reading scientific literature, and searching online to assist in molecular synthesis.

In real-world laboratory settings, general LLMs have demonstrated their effectiveness in synthesizing molecules and controlling robotic arms~\cite{boiko2023emergent, bran2023chemcrow, yoshikawa2023large}. These successes underscore the potential for general LLMs to impact the field of chemistry significantly, presenting new ways of conducting experiments and facilitating discoveries. However, despite these promising developments, there needs to be more evidence of general LLMs being deployed in the industry, e.g., pharmaceutical companies. This gap highlights the need for further research and development to extend the use of general LLMs in chemistry experiments.

\textbf{In-silico Simulation.}
The use of specialized LLMs in industry is becoming increasingly common, with tools like AlphaFold Multimer~\cite{evans2021protein} being used for protein-protein complex prediction. More recently, these tools have expanded to include protein-ligand complexes, broadening their scope to small molecules and nucleic acids. Additionally, In Silico Medicine, a biotechnology company has developed GENTRL~\cite{zhavoronkov2019deep} and Chemistry42~\cite{ivanenkov2023chemistry42}, which utilized specialized LLMs to identify potent DDR1 kinase inhibitors and generate novel molecular structures, respectively. These structures exhibit optimized properties and have been validated through extensive in vitro and in vivo studies. Similarly, Molformer~\cite{ross2022large}, developed by IBM, demonstrates promising results in generating candidate molecules aimed at inhibiting the SARS-CoV-2 virus and developing antimicrobial peptides~\cite{das_antimicrobial2021}. 

Conversely, applying general language models remains primarily confined to in-silico environments. These models face substantial challenges in scientific understanding and quantitative analysis. For instance, GPT-4~\cite{ai4science2023impact} struggles with interpreting and generating SMILES strings. Additionally, general LLMs often lack the precision required for quantitative tasks, leading to suboptimal performance in simulations, such as predicting binding affinity~\cite{razeghi-etal-2022-impact}.

\textbf{ADMET Prediction.}
For specialized LLMs, IBM's Molformer~\cite{ross2022large} has built a cloud-based platform that allows chemists to conduct real-time molecular screening and efficient molecular properties prediction.

In the advanced stage, LLM4SD~\cite{zheng2023large} uses general-purpose LLMs such as Falcon~\cite{penedo2023refinedweb} and Galacica~\cite{taylor2022galactica} as backbones to extract meaningful hypotheses from ADMET data. These assumptions have been shown to outperform state-of-the-art professional baselines when applied to traditional methods such as random forests. Pharmacologists validate the majority of these rules, ensuring their relevance and efficacy.

\textbf{Lead Optimization.}
Specialized LLMs, for lead optimization have been validated via real-world experiments. For instance, in molecular optimization, Moret et al.~\cite{moret2023leveraging} developed a chemical language model that facilitated the discovery of a new PI3K$\gamma$ ligand with sub-micromolar activity. Similarly, in protein optimization, Hie et al. ~\cite{hie2023efficient} employed a language-model-guided process to enhance seven antibodies' affinity and effectiveness against Ebola and SARS-CoV-2 through minimal variant screening and lab evolution. 

However, general LLMs are still in the early stages of development and have only undergone in-silico testing. One of the main challenges in adapting general LLMs to this application lies in the profound understanding of scientific language, which is essential for lead optimization.

\textbf{Assistance.}
General LLMs have reached an advanced stage in information retrieval and explanation for drug discovery. BenevolentAI, a company specializing in AI-enabled drug discovery and development, is investigating ChatGPT retrieval plug-ins that can search through personal or company documents to provide medical answers~\cite{savage2023drug}. Furthermore, GPT-4 showcases substantial coding capabilities; it aids in various coding tasks related to drug discovery, such as data downloading and data preprocessing~\cite{ai4science2023impact}.

\subsubsection{Clinical Trial}
The clinical trial phase mainly involves general text data, including electronic health records and trial protocol documents. Therefore, a specialized LLM is generally not suitable at this stage.

\textbf{Clinical Trial Practice.}
In clinical trial practice, there are tasks including ICD coding, patient-trial matching, and clinical trial planning. Although general LLMs in this field are still in the early stages of implementation, their potential is promising. Despite the lack of extensive real-world testing and application evidence, the rapid development and improvement of the models, particularly in their ability to understand and process medical knowledge~\cite{singhal2023large}, signals a promising future.

\textbf{Patient Outcome Prediction.}
Predicting patient outcomes is one area where the LLMs show promise, helping doctors diagnose and predict patient outcomes. General LLMs specialize in handling large amounts of unstructured data in electronic medical records. For example, Jiang et al. ~\cite{jiang2023health} developed NYUTriton, an advanced platform that interfaces with the electronic health record system at NYU Langone Health System. Deployed across a network of hospitals and outpatient facilities in New York, the system performs tasks such as predicting in-hospital mortality, estimating a comprehensive comorbidity index, and predicting 30-day all-cause readmissions. Similarly, Google's MedPalm2~\cite{singhal2023large} was introduced for medical question-answering tasks and achieved an accuracy of 86.5\%, much higher than the approximate medical passing score. This advanced technology is being tested in real-world settings with select client groups, including VHC Health VA, affiliated with Mayo Clinic.

\textbf{Assistance.}
General LLMs have matured in clinical assistance. These assistants can aid physicians and administrative staff in tasks such as document writing. For example, Webster \& Paul~\cite{webster2023six} have demonstrated the effectiveness of these models in generating clinical notes, maintaining regular check-ins for patients with chronic conditions, and summarizing patient issues. Recently, Oracle unveiled a Clinical Digital Assistant that can handle administrative tasks through voice commands. Additionally, Google's MedPalm2 has been implemented in real-world scenarios for information retrieval and knowledge explanation~\cite{singhal2023large}.

\section{Future Direction}
\label{sec:future}
This section explores the future direction of LLM applications in drug discovery and development. We discuss nine areas that require enhancement: integrating biological insights, addressing ethical and privacy concerns and preventing misuse, addressing fairness and mitigating bias, addressing hallucination, improving multi-modality, improving context window, improving spatio-temporal understanding, and integrating specialized LLMs with general LLMs.

\subsection{Integrating Biological Insights}
Improving the scientific understanding of language models (LLMs) is crucial for their successful application in drug discovery and related downstream tasks. To be practical, specialized and general LLMs require a deep understanding of scientific concepts, such as terminologies, and languages, such as SMILES and IUPAC nomenclature.

A case is molecular generation and editing that requires accurate interpretation and manipulation of these specific, highly scientific languages. Microsoft performed one such evaluation using GPT-4, showing that the model could not understand SMILES strings well~\cite{ai4science2023impact}. Similarly, in clinical trials, LLMs must be familiar with medical vocabulary from Electronic Health Records (EHRs) and patient profiles regarding diagnoses and clinical criteria. Such knowledge is necessary for reliable patient clinical trial matching, a complicated multistage process.

Benchmarking the scientific understanding of LLMs is a pivotal step in their development and deployment. Though medical QA datasets are available, they may need to be more comprehensive and practical as it is hard to discern whether the model uses scientific knowledge or reproduces answers based on memorization. Hence, more rigorous and customized methods should be used to estimate and develop LLM's scientific understanding. 

Arguably, the most applicable developments come in developing explanatory capabilities at a big-data scale in biochemistry and biophysics. These technologies primarily include high-throughput secondary structure prediction technological developments and increasingly robust statistical mechanics predictions.  

Recently, high-throughput secondary structure generation for DNA and RNA represents a significant advancement in molecular biology and bioinformatics. These technologies combine in vitro biophysical probing techniques with statistical methods and technologies to predict the secondary structures of nucleic acids at a large scale, which is crucial for understanding their function, interaction, and role in various biological processes. Developing high-throughput sequencing-based technologies has provided a wealth of genomic and transcriptomic data. Future research will increasingly integrate computational predictions with experimental data, like SHAPE \cite{Loughrey_Watters_Settle_Lucks_2014} (Selective 2'-Hydroxyl Acylation analyzed by Primer Extension) or DMS \cite{Zubradt_Gupta_Persad_Lambowitz_Weissman_Rouskin_2016} (dimethyl sulfate) mapping, to refine secondary structure models.

Recent developments in statistical mechanics have also significantly enhanced the robustness of its predictions, which have important implications for LLMs in drug discovery. In particular, enhanced sampling techniques and multiscale modeling are advanced computational strategies used to overcome limitations in traditional molecular simulations, providing more accurate and comprehensive insights into biomolecular processes. Metadynamics \cite{Bussi_Laio_2020} is a popular enhanced sampling technique that facilitates the exploration of a molecule's energy landscape more efficiently than conventional molecular dynamics simulations. Metadynamics can be applied to study the binding mechanisms and conformational changes of ligands and proteins in drug discovery. The QM/MM \cite{Böselt_Thürlemann_Riniker_2021} (quantum mechanics/molecular mechanics) approach is a prime example of multiscale modeling. In QM/MM simulations, the region of interest (e.g., a drug interacting with its binding site) is treated using quantum mechanics to model the electronic interactions accurately. In contrast, the surrounding environment (e.g., the rest of the protein and solvent) is treated using classical molecular mechanics, balancing accuracy and computational efficiency. QM/MM simulations can be particularly valuable for studying enzyme-catalyzed reactions, which are often targets in drug design. 

The integration of advanced computational techniques developed in fields like statistical mechanics or molecular dynamics into large language models (LLMs) for drug discovery has been gradual due to several factors. Despite the significant progress in both domains over the last decade, the following reasons explain why some advanced techniques have not yet been widely adopted. This is potentially due to interdisciplinary gaps, validation and standardization protocols, and data compatibility. As interdisciplinary collaboration grows and computational resources become more accessible, we will likely see more advanced techniques integrated into LLMs, enhancing their effectiveness and impact in drug discovery.

\subsection{Addresing Ethical, Privacy Concerns, \& Preventing Misuse}
The ethical issues in using LLMs for drug discovery are diverse and involve responsibility, fairness, and the potential for unintended consequences. One primary ethical issue revolves around accountability for decisions made or influenced by LLMs. As these models play an increasingly important role in drug development, the question arises of who is responsible for the outcomes, whether positive breakthroughs or negative results. This is particularly challenging given the often opaque nature of the LLM decision-making process. The rapid pace of innovation in this field has raised concerns that regulations and ethical guidelines must be updated.

Privacy issues related to LLMs are essential because they can memorize training data. For example, when it comes to sensitive multi-omic data collected during patient typing, it is critical to ensure the data is anonymised and therefore cannot be directly related to the patient. Recently, a study showed that adversaries could extract large amounts of training data from LLM~\cite{nasr2023scalable}. This is achieved through extractable storage, where the model accidentally leaks training data in response to specific queries. For example, a model can be reverted to its original language modeling behavior by prompting ChatGPT with a sequence that causes it to break from Chatbot-style generation. This difference can cause the model to output fragments of its training data. Researchers developed a new divergence attack for ChatGPT that exposed training data 150 times higher than usual. These results indicate that current alignment techniques do not entirely prevent memory leakage, which raises serious ethical and safety issues.

Potential misuse of LLMs in areas such as drug discovery necessitates a carefully balanced approach that prioritizes safeguarding against risks without impeding technological advancement. These concerns are raised as LLMs can be used intentionally for malicious purposes as described in MegaSyn2 model~\cite{urbina2022dual}. They show the MegaSyn2 model~\cite{urbina2022dual}, initially developed to discover therapeutic inhibitors and exploit them to create highly toxic substances or chemical warfare agents. However, it is crucial to acknowledge that while LLMs democratize the knowledge of compound synthesis, this does not automatically lead to more accessible access to materials for synthesizing dangerous substances due to existing strict regulations. Given the LLMs' potential benefits in areas like medical science, they need more relaxed regulations to ensure their development. This also requires a balanced view considering system safety while avoiding slowing down technology development.

\subsection{Addresing Hallucination}
The use of language large models (LLMs) in drug discovery is growing. However, their tendency to " hallucinate"—generating irrelevant or incoherent responses—presents a major challenge. Researchers and clinicians must be cautious, as these errors can lead them in the wrong direction with false information. These errors can be propagated, which leads to serious consequences. 

For example, hallucinations can potentially result in the identification of incorrect biological targets or relationships, driving research in unproductive directions and wasting valuable resources. Some biotech companies are now using LLMs to interact with knowledge graphs of biological entities, such as genes, proteins, and diseases, to identify potential targets for drug development~\cite{savage2023drug}. This issue can lead to inaccurate optimization or modifications in molecule and protein design. For instance, hallucinated molecular structures could be chemically invalid or impractical for synthesis. In clinical settings, where the stakes of inaccuracy include diagnosis and data interpretation, it can lead to serious, even life-threatening consequences.

To address hallucinations in LLMs for drug discovery, mitigation strategies can be used to guide the model toward generating more accurate and relevant answers. Knowledge editing is an approach that fills gaps in the understanding of the model by modifying some parameters or integrating plugins from other sources~\cite{ji2023survey}. This entails grounding LLMs in retrieval-augmented generation (RAG) with external documents for increased factuality and relevance in their outputs~\cite{ji2023survey}. Similarly, fine-tuning LLMs on debiased datasets helps remove knowledge shortcuts and spurious correlations that might stem from biased sampling~\cite{ji2023survey}. Additionally, many techniques can be applied to improve knowledge recall. Chain of Thoughts Prompting and similar techniques help generate outputs based on factuality and relevance~\cite{ji2023survey}. Finally, the effort of perfecting the decoding algorithms, such as Factuality Enhanced Decoding and Faithfulness Enhanced Decoding, ensures that the output generated is entirely in line with actual data and customer requests, significantly enhancing both the accuracy and reliability of LLMs.

\subsection{Addresing Fairness \& Bias}
Fairness and bias should be among the top priorities when creating and using LLMs in drug development. Biases' effects are evident in medical contexts, which can lead to possible inaccuracies, discrimination between different groups of patients, or even potentially harmful consequences. Biases inherent within data collection, model training, and application channels may prolong disparities, thus negatively impacting the integrity and efficacy of medical treatments.

Bias can originate from various sources, including the need for more data on rare diseases or specific populations, leading to underrepresentation. For example, biases are particularly evident in clinical trials through disparities in participation rates among different populations, influenced by geographic, economic, and cultural factors. Skewing predictive models developed based on these datasets can result in less effective or inappropriate solutions for underrepresented groups. Furthermore, this problem is accentuated when the models turn more toward common or widely examined points, like particular protein targets and demographic groups.

Enhancing the transparency and interpretability of LLMs is essential to address their biases. This can be achieved by employing various data sources, utilizing inclusive data collection and analysis approaches, and conducting rigorous ethical assessments. Furthermore, creating mechanisms that allow for error rectification, promoting interdisciplinary cooperation, and initiating conversations concerning the responsible implementation of LLMs in drug development would help ensure equity in healthcare access.

\subsection{Improving Quantiative Analysis}
The role of LLMs in the field of drug discovery has been expanding. One significant skill they need is to analyze vast amounts of numerical datasets. For instance, this applies to transcriptomic expression data interpretation for learning about disease mechanisms or predicting molecular properties while making drugs~\cite{theodoris2023transfer, zheng2023large}. These examples indicate a growing trend towards relying on LLMs to manage particular problems within data-intensive pharmaceutical research.

LLMs, in general, while proficient in text generation and analysis, have shown limited success with data predominantly comprising numerical values, a critical aspect in drug research. For example, LLMs have historically faced challenges executing straightforward arithmetic operations like multi-digit multiplication~\cite{dziri2023faith}. They often resort to fabricating answers~\cite{gpt4-technical-report,frieder2023mathematical}. As some have argued~\cite{testolin2023can,golkar2023xval}, this can be due to the standard tokenization methods in LLMs that fail to accurately reflect the unique quantitative characteristics of numerical data, separating it from typical language inputs.

Recent explorations propose various methodologies to improve the encoding of numerical information for these LLMs~\cite{thawani2021representing,golkar2023xval}. Approaches include digit-by-digit encoding~\cite{gruver2023large}, base-10 formatting, and alignment between embedding distance against real numerical distance~\cite{sundararaman2020methods}. However, as LLMs are prone to relying on shortcuts and non-representative correlations in data ~\cite{dziri2023faith}, they still struggle with interpolation and extrapolation in mathematical contexts within scientific fields~\cite{grosse2023studying,anil2022exploring}. Addressing this fundamental issue requires imposing an appropriate inductive bias that acknowledges the continuous nature of numbers, a critical step for advancing LLMs in drug discovery.

\subsection{Improving Multi-Modality}
Multimodal Large Language Models (MLLMs) are advanced LLMs equipped to receive and process multimodal information~\cite{yin2023survey}. The application of MLLMs to drug discovery is promising as they can process diverse types of data, including videos, images, and experimental data. This property aligns with the nature of drug discovery, requiring diversified data sources like chemical structures, biological datasets, and scientific literature~\cite{taylor2022galactica}. 

MLLMs are beneficial as they enable scientists to interact intuitively and flexibly with them. They can be accommodating when it comes to complex tasks in chemistry or biology, such as molecular modeling or clinical data analysis. These tasks involve multiple types of data, e.g., 3D protein and molecule structure and 2D medical images. 
Implementing MLLMs can open up an exciting frontier for research and may significantly optimize the effectiveness of laboratory investigations.

\subsection{Improving Context Window}
In drug discovery, LLMs usually must deal with vast amounts of biological data like sequences, which can easily exceed 2048 tokens~\cite{koh2022empirical,koh2022far}. However, many existing models have a restricted window of 2048 or 4096 tokens, e.g., Galactica~\cite{taylor2022galactica}, and Falcon.
Thus, these models are ineffective in processing multiple proteins and gene sequences. Protein sequences typically have 200-300 amino acids, while gene sequences can have over 26,000 nucleotides. Even LLMs with large window sizes of up to 128k tokens fail to thoroughly analyze such enormous volumes of input data. Typically, these models are good at interpreting the input's beginning and end but usually perform poorly for middle sections. This ``forgetting'' issue can lead to significant gaps in the analysis and interpretation of data.

Several potential solutions exist to address this challenge. One approach is to segment the input into smaller chunks and process them separately, then combine the outputs to generate a final result. To ensure no critical information is lost, intelligent segmentation strategies should be developed to understand the biological significance of different parts of the sequence. Another solution is implementing more sophisticated memory and attention mechanisms that help models better manage and utilize longer context windows. However, this research direction requires intensive computational resources.

\subsection{Improving Spatio-temporal Understanding}
An essential prerequisite for developing rational drug design and discovery is the improvement of spatial-temporal capabilities in LLMs, given that these techniques rely heavily on processing large datasets comprising complex, multi-dimensional data. At present, LLMs can process and interpret textual information reasonably well but have their weaknesses exposed when it comes to spatio-temporal data~\cite{pan2023unifying,jin2023time,jin2023large}, which plays an essential role in the drug discovery field. For example, this limitation leaves aside areas where a physical change concerning time and space understanding is crucial, e.g., dynamic interactions between molecules. 
Improving LLMs in this aspect would provide new opportunities to understand more deeply in fields like spatial-temporal transcriptomics and
molecular dynamics simulations~\cite{nguyen2024application}.

Furthermore, LLMs with enhanced spatial-temporal and multi-modal understanding enable a highly autonomous and efficient process. An illustration is in the analysis of molecular dynamics simulations. These models can automatically investigate, document, and even describe data that elaborate on molecule dynamics. This advancement and multi-modal capabilities are pivotal in unearthing potential drug candidates and uncovering molecular pathways commonly hidden beneath vast data.
This exciting development has the potential to revolutionize drug discovery and significantly reduce time and resource expenditure.

\subsection{Integrating Specialised LLMs \& General LLMs}
Combining a specialized language model and a general-purpose LLM gives an edge in drug discovery. Specialized LLMs perform admirably in precision tasks like understanding biological information, estimating molecule interactions, or examining protein configurations using their niche training datasets. On the other hand, general LLMs provide versatility and a broad knowledge base that can be applied to various subjects and tasks. They are considered essential instruments for many users since they are user-friendly to individuals with different levels of professionalism, such as researchers and medical workers, enabling them to access scientific knowledge and reasoning easily.

Specifically, general LLMs can act as the front-end system responsible for user interaction with conversational interfaces that provide detailed descriptions of case characteristics, assist in problem identification, and facilitate discussion of decision alternatives. They might deliver context, background knowledge, and reasoning ability to aid in understanding the problem situation. On the other hand, specialized LLMs with quantitative analytical ability can further be utilized to accomplish specific downstream tasks. For instance, they can conduct QSAR analysis for molecule compounds, protein folding simulation, or molecular structure optimization.
Furthermore, these LLMs can then provide the results back to general-purpose LLMs to synthesize and interpret these results and provide insightful information for users.

\nocite{langley00}

\bibliography{example_paper}
\bibliographystyle{icml2023}

\newpage


\end{document}